\documentclass[
 reprint,amsmath,amssymb,aps,superscriptaddress,
]{revtex4-2}
\usepackage{graphicx} 
\setcitestyle{super}
\usepackage{physics}
\usepackage{float}
\usepackage{multibib}
\usepackage{lineno}
\usepackage{xcolor}
\usepackage{braket}
\usepackage{textcomp}

\newcommand{\parens}[1]{\left(#1\right)}
\newcommand{\sparens}[1]{\left[ #1 \right]}

\usepackage{amssymb}
\usepackage{mathtools}% Loads amsmath

\usepackage{times}
\usepackage{float}
\newcommand{\dirac}[1]{\langle#1\rangle}

\newcites{SI}{Supplementary}

\begin{document}

\title{
Observation of Tunable Superradiant Frequency Combs
}

\author{Tian Xie}
    \thanks{Current address: Department of Applied Physics and Ginzton Laboratory, Stanford University, Stanford, CA, USA}
	\affiliation{Thomas J. Watson, Sr, Laboratory of Applied Physics, California Institute of Technology, Pasadena, CA, USA.}
    \affiliation{Kavli Nanoscience Institute, California Institute of Technology, Pasadena, CA, USA}
	\affiliation{Institute for Quantum Information and Matter, California Institute of Technology, Pasadena, CA, USA}
\author{Rikuto Fukumori}
	\affiliation{Thomas J. Watson, Sr, Laboratory of Applied Physics, California Institute of Technology, Pasadena, CA, USA.}
    \affiliation{Kavli Nanoscience Institute, California Institute of Technology, Pasadena, CA, USA}
	\affiliation{Institute for Quantum Information and Matter, California Institute of Technology, Pasadena, CA, USA}
\author{Wai-Keong Mok}
	\affiliation{Institute for Quantum Information and Matter, California Institute of Technology, Pasadena, CA, USA}
 \author{Jiahui Li}
	\affiliation{Thomas J. Watson, Sr, Laboratory of Applied Physics, California Institute of Technology, Pasadena, CA, USA.}
    \affiliation{Kavli Nanoscience Institute, California Institute of Technology, Pasadena, CA, USA}
	\affiliation{Institute for Quantum Information and Matter, California Institute of Technology, Pasadena, CA, USA}
    
\author{Joonhee Choi}
    \thanks{joonhee.choi@stanford.edu}
    \affiliation{Department of Electrical Engineering, Stanford University, Stanford, CA, USA}
\author{Andrei Faraon}
    \thanks{faraon@caltech.edu}
	\affiliation{Thomas J. Watson, Sr, Laboratory of Applied Physics, California Institute of Technology, Pasadena, CA, USA.}
    \affiliation{Kavli Nanoscience Institute, California Institute of Technology, Pasadena, CA, USA}	\affiliation{Institute for Quantum Information and Matter, California Institute of Technology, Pasadena, CA, USA}
    
\begin{abstract}

Cavity quantum electrodynamics (QED) with quantum emitters coupled to resonators provides a powerful platform for engineering light-matter interactions and exploring collective phenomena. In particular, superradiance, arising from collective quantum interference among emitters, has been explored as a route to ultrastable continuous radiation. However, engineering superradiance in the time domain to realize periodic pulsed sources or frequency combs remains largely unexplored. Here, we investigate the non-equilibrium many-body dynamics of a driven spin ensemble coupled to an on-chip superconducting resonator and uncover a dynamical phase transition from continuous-wave to periodic pulsed superradiant emission. To quantitatively capture the observed dynamical phases, we introduce a driven-dissipative cavity-QED model that elucidates how the periodic pulsed superradiant phase emerges from collective, periodically repeating spin dynamics stabilized by the interplay of coherence growth, disorder, and dissipation. We also find that rare-earth ion spin systems exhibiting both optical and microwave transitions enable phase-synchronized, dual-rail superradiant frequency combs in both the microwave and optical domains. Our results not only open new avenues for dual-rail frequency-comb applications in quantum metrology and information processing, but also establish a fundamental connection between periodic pulsed superradiance and the emergence of a continuous time crystal as a novel nonequilibrium phase in driven open systems. 

\end{abstract}

\maketitle

\section*{Introduction}
The invention of the laser stands among the most transformative technological advances of the twentieth century \cite{schawlow1958infrared}. Laser operation relies on stimulated emission, whereby coherent radiation emerges once the gain exceeds loss in a resonant cavity \cite{maiman1960stimulated}. In contrast, superradiance (SR), first proposed by Dicke in 1954 \cite{dicke1954coherence}, can also lead to continuous coherent radiation arising from a collective many-body dissipation process in which emission results from quantum interference among an ensemble of emitters \cite{gross1982superradiance}. Remarkably, compared with conventional lasers, superradiant emission can exhibit exceptional frequency stability due to the long coherence of atomic transitions, along with output power that scales nonlinearly with ensemble size \cite{bohnet2012steady}. Superradiance has been observed across a wide range of platforms in both free space and cavity-coupled systems, including cold atoms \cite{norcia2016superradiance,kim2018coherent,supercold2016}, atomic gases \cite{skribanowitz1973observation,gross1976observation}, artificial atoms \cite{scheibner2007superradiance,lambert2016superradiance,mlynek2014observation}, and solid-state defects \cite{lei2023many,lukin2023two,bradac2017room}.

Specifically, from a conceptual perspective, conventional lasers and superradiant light sources represent two complementary limits of light–matter interaction in a cavity-coupled system. Conventional lasers rely on stimulated emission, where the coherence is governed primarily by the cavity field. Superradiant systems, however, operate in the opposite regime, where coherence resides predominantly in the atomic ensemble, and radiation emerges from collectively synchronized dissipation driven by atom–atom correlations mediated by a lossy (bad) cavity (Fig.~1a-c). In this regime, the dynamics are naturally described within the framework of cavity quantum electrodynamics, where light–matter interactions are treated at the fundamental quantum-mechanical level. This perspective has motivated extensive studies on superradiant dynamics in cavity-QED systems~\cite{walther2006cavity,reiserer2015cavity,niemczyk2010circuit,song2025dissipation}, including superradiant bursts\cite{angerer2018superradiant,kersten2026self}, triggered superradiance\cite{kersten2023triggered}, and ultranarrow-linewidth superradiant lasers \cite{bohnet2012steady,bohnet2012relaxation,norcia2018cavity}.

Historically, advances in laser physics have demonstrated that introducing nonlinear elements can drive systems into unstable dynamical regimes \cite{abraham1988dynamical}, enabling the development of mode-locked lasers that generate periodic, ultrafast optical pulses \cite{haus2002mode}. This periodic emission in the time domain, corresponding to a frequency comb in the spectral domain, dramatically expands the impact of lasers across science and industry \cite{udem1999absolute,denk1990two}. This raises a natural question: can \textit{superradiant} systems support frequency-comb-like emission that produces time-periodic, phase-coherent pulse trains, and if so, what mechanism underlies this behavior, and can the resulting time-domain pulsing be engineered?

\begin{figure*}
\centering
\includegraphics[width=1\linewidth]{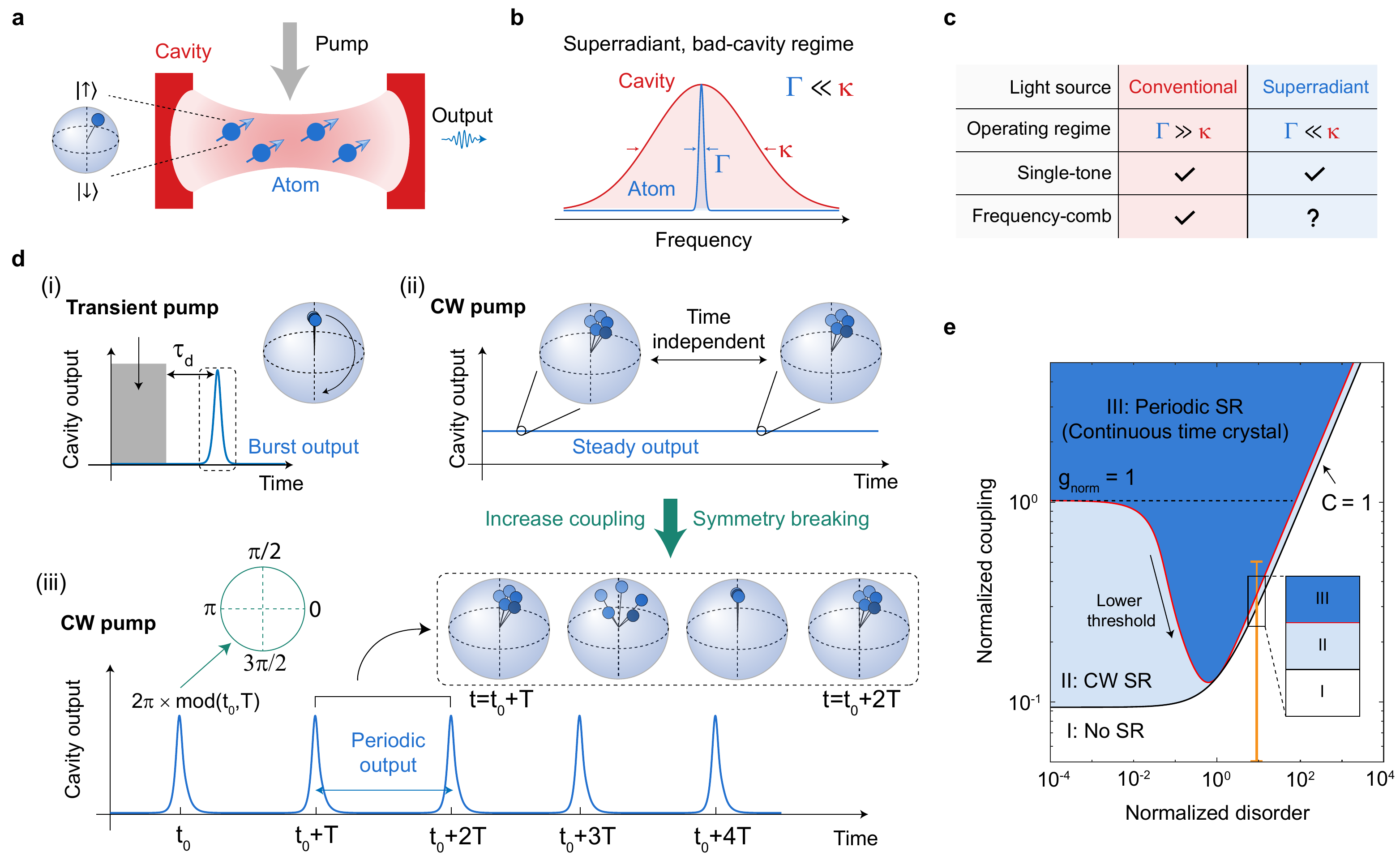}
\caption{\label{Fig1}
\textbf{Overview of superradiant dynamics in a cavity-coupled atomic ensemble. }
\textbf{a}, An ensemble of two-level atoms, $\{\ket{\uparrow}, \ket{\downarrow}\}$, coupled to a common cavity field. The cavity output serves simultaneously as a probe of the intracavity dynamics and as an emitted light field generated by the atomic ensemble. \textbf{b}, The frequency-domain picture for superradiant systems, where the inhomogeneously broadened atomic linewidth, $\Gamma$, is narrower than the cavity linewidth, $\kappa$.
\textbf{c}, Comparison table between conventional and superradiant light sources.
\textbf{d}, Three types of superradiant dynamics: (i) superradiant burst emission under a transient pump, (ii) steady superradiant emission under a CW pump, and (iii) periodic pulsed superradiant emission under a CW pump. In (iii), as the cavity coupling strength increases, periodic pulsed emission dynamics emerge due to the spontaneous breaking of continuous time-translation symmetry under a CW drive. We find that the periodicity, $T$, of the pulsed emission is set by the ensemble cooperativity, while the onset time, $t_0$, of the resulting pulse train is uniformly random. 
\textbf{e}, Phase diagram of a cavity-QED model showing three distinct phases of (I) no SR, (II) CW SR, and (III) periodic SR. These phases are distinguished by the interplay of cavity-mediated collective coupling and dissipation, and disorder due to the frequency inhomogeneity of the emitters. 
The CW SR phase emerges when the ensemble cooperativity exceeds unity ($C = 1$, black solid line), while the periodic SR phase is expected to emerge when the normalized cavity-ensemble coupling strength exceeds unity ($g_\text{norm} = 1$, black dashed line). Surprisingly, the phase transition into the periodic SR phase occurs at a weaker coupling strength in our experiment due to the presence of disorder, consistent with our theoretical prediction (red solid line). The yellow vertical bar indicates the experimentally probed regime of collective coupling strengths. See the main text and Supplementary Information for details.
}
\end{figure*}

While recent theoretical~\cite{patra2019driven} and experimental~\cite{hara2024periodic,hara2024analytical} efforts have begun to explore periodic pulsed superradiant dynamics in driven cavity QED systems, a clear understanding of their physical origin and underlying mechanisms, as well as comprehensive experimental demonstrations, remains elusive. In this work, we develop a unified theoretical and experimental framework that elucidates how superradiant frequency combs emerge in a driven cavity-QED system of quantum emitters through the interplay of atomic coherence, dissipation, and disorder, where the nonlinearity is provided by superradiant emission. Building on this framework, we experimentally investigate non-equilibrium superradiant dynamics in a disordered spin ensemble coupled to a superconducting microwave cavity, and show that robust, phase-coherent superradiant pulse trains can be both probed and engineered, with their pulse repetition rate set by the ensemble cooperativity. Unexpectedly, we also observe, for the first time, simultaneous and correlated periodic microwave and optical superradiant emissions that are both time-synchronized and phase-coherent. We refer to this phenomenon as \textit{dual-rail superradiant frequency combs}, which opens new opportunities for ensemble-based quantum technologies. Furthermore, our results reveal a fundamental connection to a continuous time crystal---a dynamical phase of matter in driven, non-equilibrium cavity QED systems characterized by spontaneously broken continuous time-translation symmetry and the emergence of robust temporal order.

% \section{Results}
\section*{Physics of superradiant dynamics}
To develop an intuitive understanding of the physical origin and mechanisms underlying superradiant frequency-comb formation, we consider an ensemble of frequency-inhomogeneous two-level quantum emitters ($\ket{\downarrow}, \ket{\uparrow}$) coupled to a common lossy cavity mode and driven by an external pump. The microscopic dynamics of such cavity-ensemble interactions are imprinted on the cavity output, which serves as the primary experimentally accessible observable of the system's cavity QED behavior. Specifically, the photon emission rate, $\Gamma_\text{ph}$, is governed by three distinct emission processes: Purcell-enhanced spontaneous emission, stimulated emission, and superradiant emission~\cite{zollitsch2025quantum}:
\begin{equation}
    \Gamma_\text{ph} =\frac{4g^2}{\kappa} 
    \Bigg[\underbrace{\left(\frac{N}{2} - \expval{\hat{S_z}}  \right)}_\text{Spontaneous}
    -
    \underbrace{2\expval{\hat{S_z}}  \expval{\hat{b}^\dagger \hat{b}}}_\text{Stimulated} 
    + \underbrace{ \expval{ \hat{S}_{\uparrow\downarrow}\hat{S}_{\downarrow\uparrow}}}_\text{Superradiant}\Bigg]
    \label{eq:photon-emission}.
\end{equation}
Here $g$ is the single atom-cavity coupling strength, $\kappa$ is the cavity decay rate, $N$ is the number of quantum emitters, $\hat{S}_z= \frac{1}{2}\sum_j\hat{\sigma}_{z,j}$ is the collective spin operator along the $z$ direction, representing the total population imbalance of the ensemble with $\hat{\sigma}_{z,j} = \ket{\downarrow}_j \bra{\downarrow} - \ket{\uparrow}_j\bra{\uparrow}$ for the $j$-th spin, $\hat{S}_{\downarrow\uparrow} = \sum_j \hat{\sigma}_{\downarrow\uparrow,j}$ and $\hat{S}_{\uparrow\downarrow} = \hat{S}_{\downarrow\uparrow}^\dagger$ are the collective ensemble coherence obtained from the sum of individual atomic coherences with $\hat{\sigma}_{\downarrow\uparrow, j} = \ket{\downarrow}_j\bra{\uparrow}$, and $\hat{b}$ and $\hat{b}^{\dagger}$ are the cavity annihilation and creation operators, respectively (Methods). 

In contrast to conventional laser systems, where emission is dominated by stimulated emission (second term in Eq.~(\ref{eq:photon-emission})), superradiant systems are governed by collective spin correlations (third term), with the photon emission rate scaling as $\langle \hat{S}_{\uparrow\downarrow}\hat{S}_{\downarrow\uparrow} \rangle \propto N^2$. This collective behavior acts as an effective nonlinearity, providing the foundation for dynamical phase transitions in superradiance.

While three different mechanisms contribute to the photon emission rate, the presence of superradiant dynamics can first be identified by monitoring the emission following a transient excitation pump pulse (Fig.~1d (i)). In this case, superradiant ensembles exhibit a delayed burst of emission arising from stochastic quantum fluctuations that initiate synchronized dissipation of individual emitters, providing a clear signature of superradiance~\cite{angerer2018superradiant,norcia2016superradiance}. 

Transitioning from a transient pump to a CW pump, the superradiant ensemble can support continuous emission by reaching a steady-state condition under the CW pump (Fig.~1d (ii)). Analogous to a conventional laser, this condition is achieved when the cavity-mediated gain, $\frac{4Ng^2}{\kappa}$, is greater than the atomic dephasing rate, $\Gamma$ (Here, $\Gamma$ is the linewidth of the emitter ensemble). Equivalently, defining the ensemble cooperativity as $C = \frac{4Ng^2}{\kappa \Gamma}$, CW superradiant emission occurs when $C \geq 1$ (Fig.~1e, black solid line).

Crucially, when the coherent cavity-ensemble coupling strength, $\sqrt{N}g$, exceeds the cavity-induced decoherence rate, $\kappa/2$, the steady-state dynamics can break down, turning CW emission into periodic pulsed emission (Fig.~1d (iii) and the black dotted line in Fig.~1e). In this regime, quantum interference between the strongly dressed cavity-ensemble states drives oscillatory behavior in the photon emission. Surprisingly, however, we find that the presence of moderate disorder (defined as an additional inhomogeneous contribution to spectral line broadening beyond the homogeneous linewidth) substantially lowers the critical coupling threshold, thereby facilitating the phase transition into the pulsed phase (red solid line, Fig.~1e). Details of the theoretical phase boundary estimate based on a Hopf bifurcation model are presented in Supplementary Information.

Using our driven-dissipative cavity QED model, we elucidate how such periodic pulsed emission directly originates from an emergent, time-periodic trajectory of the spin ensemble, in which collective coherence builds up, collapses through synchronized dissipation, and is then rebuilt under continuous pumping (inset of Fig.~1d (iii)). We find from both our experiment and the theoretical model that this persistent pulsed emission repeats robustly without decay, with a periodicity, $T \propto 1/C$, set by the ensemble cooperativity, $C$. This periodic superradiance phase, arising \textit{intrinsically} and spontaneously from time-independent cavity QED dynamics, stands in stark contrast to conventional mode-locked lasers, where the periodicity is \textit{externally} imposed by the cavity round-trip time. Such spontaneously broken time-translation symmetry, leading to the emergence of temporal order, defines a continuous time crystal, a class of non-equilibrium phases of matter that has been explored in a variety of driven–dissipative systems \cite{kongkhambut2022observation,wu2024dissipative,greilich2024robust,tucker2018shattered}.

\begin{figure*}
\centering
\includegraphics[width=1\linewidth]{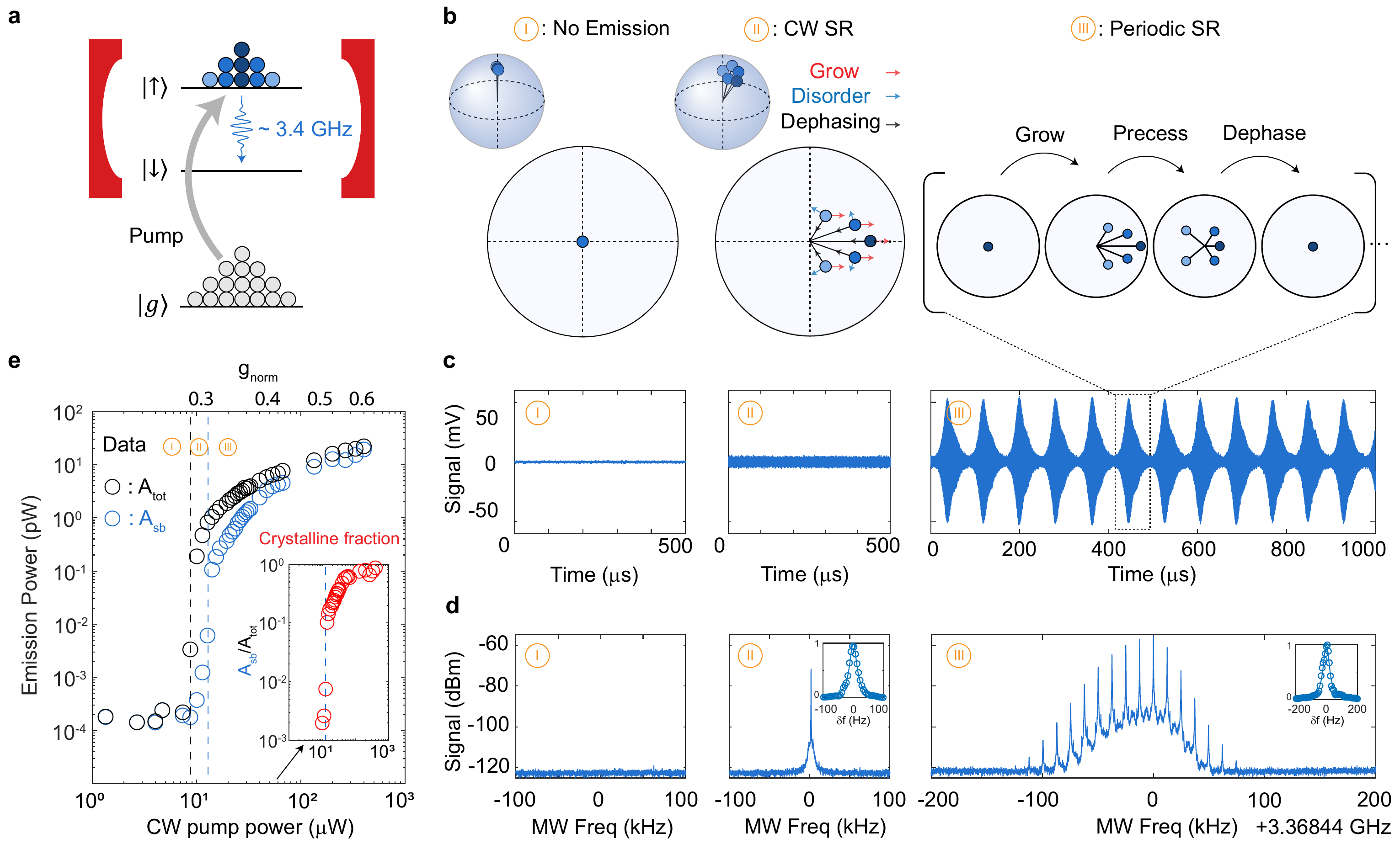}
\caption{\label{Fig2}
\textbf{Measurement of superradiant dynamics in a disordered spin ensemble.}
\textbf{a}, Schematic of the effective three-level structure, $\{ \ket{\uparrow}, \ket{\downarrow}, \ket{g} \}$, of our spin ensemble. A CW optical pump controls the net population in the two-level microwave spin manifold, $\{\ket{\downarrow}, \ket{\uparrow} \}$, by driving optical transitions between $\ket{g}$ and $\ket{\uparrow}$. Thus, the collective coupling strength, which scales with $N$, can be readily tuned by varying the pump power. A superconducting microwave resonator (red) couples the $\ket{\downarrow} \leftrightarrow \ket{\uparrow}$ transition at $\approx$3.4 GHz.
\textbf{b}, Bloch-sphere representations of superradiant dynamics in three distinct regimes (I: no SR, II: CW SR, and III: periodic SR) probed with increasing collective coupling strength. In Regime I, the collective coupling is too weak to generate ensemble coherence, resulting in no SR. In Regime II, a stable balance among cavity-induced coherence growth (red arrow), disorder-induced precession (blue arrow), and dephasing (black arrow) stabilizes CW superradiant emission for all individual spins. Five spins are depicted representatively. In 
Regime III, the CW SR phase becomes unstable due to increased coupling strength, driving the spin system into a phase characterized by persistent, periodic pulsed emission dynamics. The periodic behavior originates from a stable limit cycle, characterized by recurring cycles of coherence growth, precession, and dephasing. 
\textbf{c}, Time-domain superradiance emission profiles for the three regimes, as the CW pump power increases from 0 $\mu$W to 36 $\mu$W. See Methods for measurement details.
\textbf{d}, Corresponding Fourier spectra for the three regimes. Insets in Regimes II and III show linewidths of $\approx$30 and $\approx$50 Hz, respectively, three orders of magnitude narrower than the microwave inhomogeneous linewidth ($\approx$160 kHz).
\textbf{e}, Microwave superradiant emission power as a function of CW pump strength. The corresponding normalized collective coupling strength, $g_\text{norm}$, is shown on the top $x$-axis (see yellow line in Fig.~1e and Extended Data Fig.~1). Black (blue) points denote the total (sideband) spectral power, $A_\text{tot}$ ($A_\text{sb}$). The phase transitions into the CW and periodic phases are marked by sharp increases in $A_\text{tot}$ and $A_\text{sb}$, respectively. Inset shows the time-crystalline fraction, $A_{\text{sb}}/A_{\text{tot}}$, providing a clear distinction between the CW and periodic SR phases as a function of pump power.
}
\end{figure*}

\section*{Experimental results}
To experimentally explore superradiant dynamics, we employ rare-earth-ion–doped crystals, which offer highly coherent transitions in both the microwave and optical domains at
cryogenic temperatures \cite{zhong2015optically,wang2025nuclear}, and are readily compatible with on-chip cavity-QED architectures \cite{ourari2023indistinguishable,zhong2017nanophotonic,wang2022high,dold2019high}. In particular, we use an ensemble of ytterbium-171 ions doped in a yttrium orthovanadate crystal \cite{kindem2018characterization}, whose excited-state spin manifold forms a two-level system, $\{ \ket{\uparrow}, \ket{\downarrow} \}$, with a microwave transition frequency of $\omega=2\pi\times3.4$ GHz and an inhomogeneously broadened linewidth of $\Gamma=2\pi\times160$ kHz (Fig.~2a). A superconducting microwave resonator with a cavity linewidth of $\kappa=2\pi\times3.6$ MHz is fabricated on the crystal to couple to the excited-state microwave spin transition. We measured a maximum collective coupling strength of $\sqrt{N}g \approx 2\pi \times 1.1$ MHz, corresponding to a normalized coupling strength of $g_\text{norm} \approx \mathrm{0.6}$, which lies below the critical threshold necessary for inducing the periodic superradiant phase in the absence of disorder (Extended Data Fig.~1). 

Importantly, although the total available number of Yb ions in the cavity is on the order of $N_{\text{tot}} \approx 4 \times 10^{10}$, we can control the effective emitter number, $N \le N_{\text{tot}}$, that participates in superradiant dynamics by varying the optical pump power that transfers population to the $\ket{\uparrow}$ state in the excited-state manifold (Fig.~2a). We experimentally verify that superradiance dominates the dynamics of our dense cavity-coupled spin ensemble by observing that the delay time, $\tau_d$, of a superradiant burst under a transient pump scales inversely with the effective ensemble size, $N$, i.e., $\tau_d \propto 1/N$ (Extended Data Fig.~2). Notably, by simply varying the CW pump power, we can continuously tune the ensemble cooperativity, thereby enabling systematic exploration of superradiant dynamics and phase transitions across three distinct regimes: no SR, CW SR, and periodic SR, which we denote as Regimes I, II, and III, respectively (Fig.~2b; Fig.~1e yellow line).

We first probe non-equilibrium dynamics under a weak CW pump at low collective cavity coupling strength (Regime I). In this regime, the $\ket{\uparrow}$-state population is insufficient to establish collective behavior, and no superradiant emission occurs (Fig.~2c,d I). 

However, as the pump power increases and crosses the critical threshold at which the ensemble cooperatively exceeds unity, a steady-state coherence emerges in this non-equilibrium setting, giving rise to CW superradiance (Regime II). Intuitively, in the Bloch-sphere picture, this can be understood as a balance of competing contributions acting on each spin, where disorder, dephasing, and coherence growth terms compensate one another, resulting in a collectively enhanced steady-state coherence  (Fig.~2b, Fig.~2c II). In the Fourier spectrum, we find that the emission linewidth is $\approx$30 Hz, orders of magnitude narrower than the inhomogeneous ensemble linewidth (Fig.~2d II, Extended Data Fig.~3).

Upon further increasing the pump power, and thus the collective coupling strength, we observe that the CW superradiant emission becomes unstable, leading to a breakdown of the balance, which in turn gives rise to time-periodic pulsed emission dynamics (Regime III). Essentially, a stronger pump provides a larger nonlinear perturbation in the collective ensemble coherence, which triggers synchronized dissipation across the entire spin ensemble. The decayed emitters are then repumped into the excited state, forming a closed, repeating cycle that generates a persistent superradiant pulse train (Fig.~2c III). This behavior can also be understood as a Hopf bifurcation that signals the breakdown of steady-state dynamics and leads to the emergence of a stable limit cycle \cite{kongkhambut2022observation}. In the spectral domain, we observe a frequency-comb structure with a narrow comb linewidth of $\approx$50~Hz, indicating stable phase coherence of the pulses (Fig.~2d III).

To systematically determine the phase boundaries across the three distinct superradiant phases, we analyze the emission spectrum in detail (Fig.~2e). Specifically, we define the total emitted spectral power as $A_\text{tot} = \int_{-\infty}^{+\infty} S(\omega) d\omega $, and the total sideband power as $A_\text{sb} = \int_{-\infty}^{\omega_p - \Delta \omega/2 } S(\omega) d\omega + \int_{\omega_p +\Delta \omega/2}^{+\infty} S(\omega)d\omega$, where $S(\omega)$ is the power spectral density of the time-domain emission signal, $\omega_p$ is the frequency corresponding to the peak spectral density and $\Delta \omega$ is a narrow integration window that isolates the central spectral peak. These quantities serve as effective experimental order parameters that distinguish the three phases in our experiment. Concretely, the onset of CW SR (Regime I to II) is marked by a sharp increase in $A_\text{tot}$, whereas the transition from the CW to the pulsed SR phase (Regime II to III) is signaled by a sharp increase in $A_\text{sb}$. To more clearly characterize the second phase transition, we introduce the normalized sideband fraction, defined as the ratio $A_\text{sb}/A_\text{tot}$. This quantity, which has also been used to characterize the crystalline fraction in time-crystal phenomena~\cite{kongkhambut2022observation}, provides a clear signature of the transition into the periodic SR phase (Fig.~2e inset). 

Crucially, this robust temporal ordering in the superradiant dynamics emerges \textit{spontaneously} under strictly time-independent CW pumping, without any externally imposed modulation or timing reference. By repeating the experiment under identical conditions, we find that the onset time of the superradiant pulse train is uniformly random from experiment to experiment (Extended Data Fig.~4), providing direct evidence of the spontaneous breaking of continuous time-translation symmetry~\cite{kongkhambut2022observation}. Our experimental measurements and numerical simulations consistently indicate that the system satisfies the defining criteria of a time-crystalline phase (Extended Data Table~\ref{CTC_table}).

\begin{figure}
\centering
\includegraphics[width=1\linewidth]{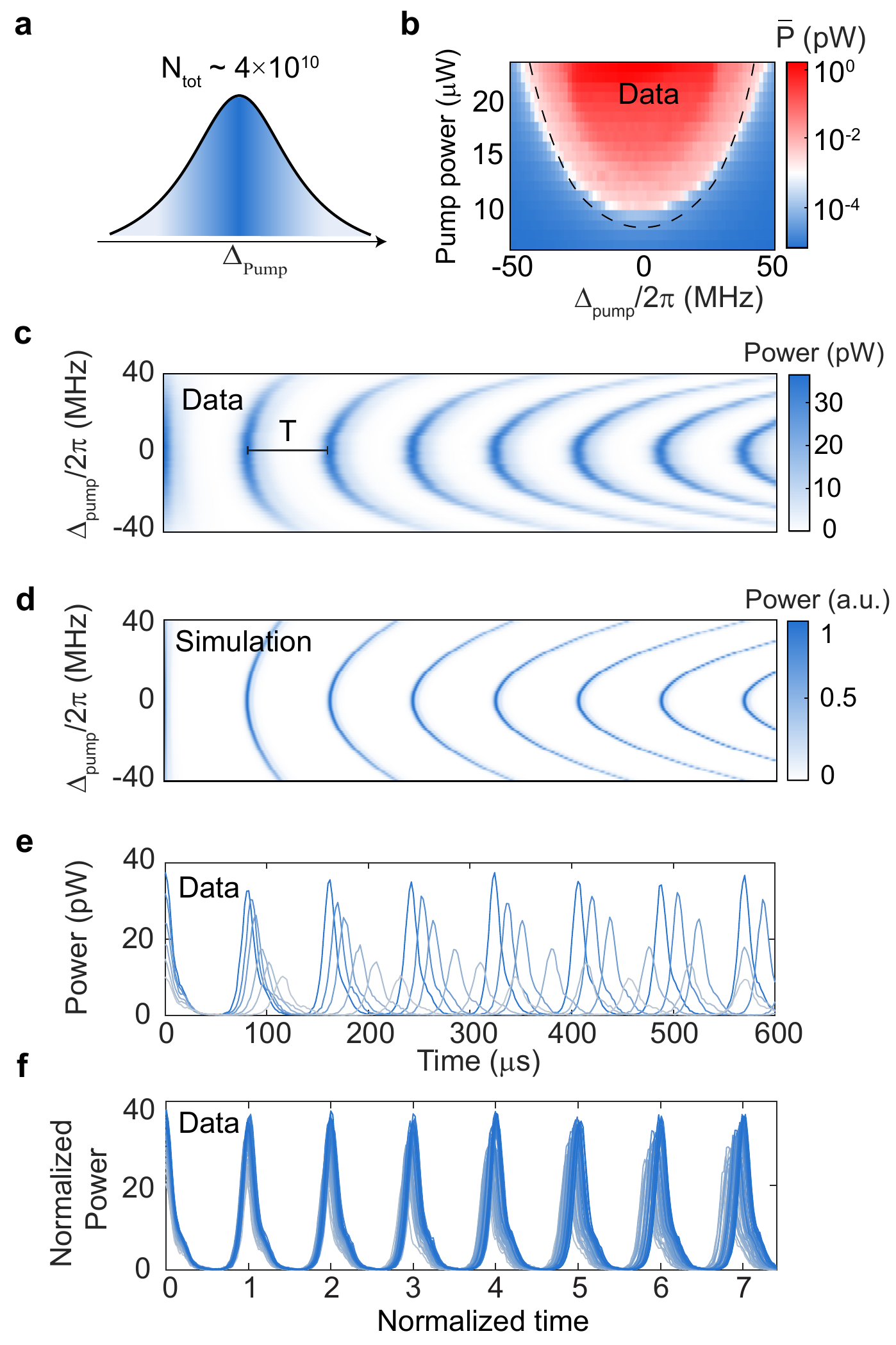}
\caption{\label{Fig3}
\textbf{Origin of periodic pulsed superradiance.} 
\textbf{a}, Optical inhomogeneous broadening profile with a linewidth of $\approx 2\pi\times92$ MHz for $N_\text{tot} \approx 4 \times 10^{10}$ ions in our system. By tuning the CW pump detuning, $\Delta_\text{pump}$, relative to the ensemble center frequency, different spectral subsets of spins are selectively pumped into the microwave exited-state manifold, enabling continuous control of the effective ensemble size that participates in superradiance.
\textbf{b}, Measured phase diagram showing superradiant emission (red, Regimes II/III) and no SR (blue, Regime I) phases under varying pump detuning and power. The dashed line indicates the calculated phase boundary corresponding to an ensemble cooperativity of 1.
\textbf{c}, Time-domain microwave SR bursts recorded at various pump detunings under 36 $\mu$W pump power. The SR burst periodicity, $T$, varies systematically with pump detuning.
\textbf{d}, Numerical simulations of the periodic SR phase based on the same experimental conditions used in (\textbf{c}), showing good agreement between theory and experiment.  
\textbf{e}, Detuning-resolved experimental time traces from (\textbf{c}) at pump detunings of $\Delta_\text{pump}/2\pi$ = 2.4, 12, 18.4, 26.4, 31.2, 36 MHz, illustrating how periodicity and emission amplitude vary with detuning. Darker (fainter) colors correspond to smaller (larger) detuning values.
\textbf{f}, Rescaled time traces from (\textbf{e}), demonstrating universal data collapse. See the main text for details.
}
\end{figure}

\section*{Origin of periodic superradiance}
A natural follow-up question is: what sets the periodicity of the persistent, periodic SR phase? According to our theoretical framework, the periodicity of the superradiant bursts is determined by the ensemble cooperativity, which depends on the number of emitters participating in the superradiant dynamics. To probe this experimentally, we selectively excite different spectral subsets of Yb ions by intentionally detuning the pump carrier frequency across the optical inhomogeneous linewidth of $\approx$92~MHz. This allows us to systematically vary the number of emitters excited to the $\ket{\uparrow}$ state within the microwave superradiant manifold while keeping the pump Rabi frequency unchanged (Fig.~3a). 

\begin{figure*}
\centering
\includegraphics[width=1\linewidth]{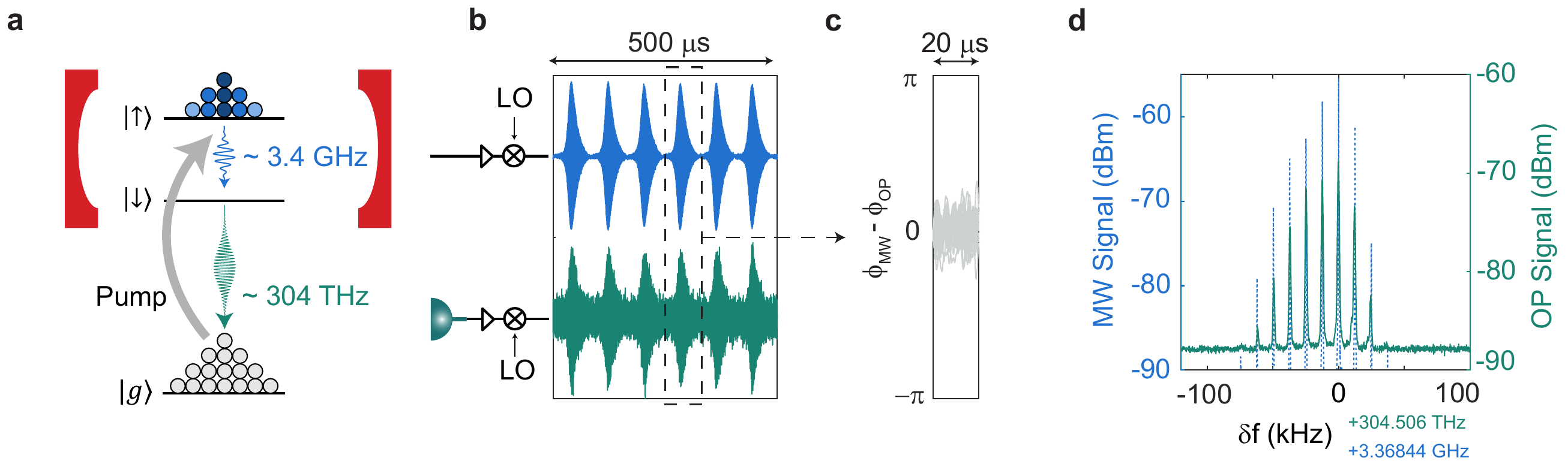}
\caption{\label{Fig4}
\textbf{Observation of dual-rail, phase-synchronized microwave and optical frequency combs.}
\textbf{a}, Transferred optical superradiance at 304 THz (984 nm) generated alongside microwave superradiance at 3.4 GHz. Under CW optical pumping, which transfers population from $\ket{g}$ to $\ket{\uparrow}$, microwave-domain spin coherence is phase-coherently transferred to the optical domain, and the resulting optical superradiance is emitted through a weak optical cavity.
\textbf{b}, Simultaneously recorded microwave and optical superradiant burst emissions under 200 $\mu$W pump power over a 500 $\mu$s window, showing time-synchronized pulsed superradiance emission. Mixers with a local oscillator (LO) are used for measurement (Supplementary Information). See Extended Data Fig.~7 for measurements taken over a 100 ms window time.
\textbf{c}, Relative phase difference between microwave ($\phi_\text{MW}$) and optical ($\phi_\text{OP}$) superradiant bursts. Individual pulses are analyzed via IQ quadratures using a 20 $\mu$s zoom-in window on each burst within a 2 ms continuous measurement frame, revealing strong in-phase correlation, i.e., $\phi_\text{MW} - \phi_\text{OP} \approx 0$ with a standard deviation of $\approx$ 0.2, across the two emission channels.
\textbf{d}, Emission spectrum of the transferred optical superradiance (green), overlaid with that of the microwave superradiance (blue, identical to that shown in Fig.~2d), revealing a spectrally overlapping frequency-comb structure.
}
\end{figure*}

As shown in Fig.~3b, the threshold for the onset of superradiant emission increases as the optical pump is detuned away from the center of the inhomogeneous linewidth, indicating that pump detuning serves as an effective control knob for tuning the collective ensemble–cavity coupling strength. In addition, we observe that in Regime III, the superradiant burst periodicity, $T$, also depends sensitively on the pump detuning (Fig.~3c): $T$ increases as the detuning moves away from zero. Numerical simulations show strong quantitative agreement with our experimental observations, further corroborating the effective tuning of the collective coupling strength via pump detuning in our system (Fig.~3d, Extended Data Fig.~5). 

Since our theoretical framework reveals that the periodicity, $T$, scales inversely with the ensemble size $N$, while the amplitude scales quadratically with $N$, we rescale both the time and amplitude axes of all measured superradiant burst profiles, acquired at different detuning values, by their corresponding effective ensemble size. Remarkably, we find that all data collapse onto a single universal curve (Figs.~3e,f). This universal collapse demonstrates that the periodic SR phase is governed by collective scaling laws set by the effective size of the driven spin ensemble, independent of the microscopic details of individual emitters.

Collectively, our experimental findings, in agreement with our theoretical model, indicate that the periodic SR phase is a robust collective phenomenon representing a non-equilibrium phase of matter in driven open quantum systems. To further substantiate this, we perform extensive numerical simulations addressing the following three key aspects (Extended Data Fig.~6). First, we confirm that, within our experimental regime, the nonlinear superradiant term associated with collective ensemble correlations dominates the conventional stimulated emission term by more than an order of magnitude. Second, we verify that the observed pulsed phase originates from a large atomic ensemble, as it cannot be explained by a single emitter with an artificially enhanced coupling strength. Third, we demonstrate that disorder plays an essential role, as the periodic superradiant phase does not emerge in the absence of disorder for the experimentally measured coupling strength, $g_\text{norm} \approx 0.6$. This finding implies that inhomogeneous broadening is not merely a source of decoherence, but can instead be harnessed to engineer collective quantum interference, thereby reshaping photon-emission dynamics in the cavity-QED system. 

\section*{Dual-rail frequency combs}
While our experimental results thus far have focused on superradiance in the microwave domain, our Yb ion-based system intrinsically comprises a three-level structure involving both microwave and optical transitions. We find that the periodic superradiant dynamics in the microwave transition are naturally transferred to the optical transition, giving rise to correlated, dual-rail frequency-comb generation in both the microwave and optical domains (Fig.~4a, Extended Data Fig.~7). As shown in Fig.~4b, the optical and microwave signals exhibit temporally synchronized burst sequences, demonstrating that both emissions originate from the same underlying collective dynamics. To quantify the coherence between the two emission channels, we analyze their relative phase difference and observe a zero-phase offset, indicating in-phase locking across the microwave and optical domains (Fig.~4c). We further characterize the spectra of the dual-rail frequency combs and confirm that the ultranarrow comb structures are well matched and exhibit the same repetition rate (Fig.~4d).

\section*{Discussion and outlook}
Our work positions disordered solid-state cavity-QED systems as a versatile platform at the intersection of cavity QED, non-equilibrium physics, and quantum metrology. A central result is a unified understanding of dynamical phase transitions in a driven cavity-QED system, where the periodic pulsed superradiant phase can be identified as a continuous time-crystal phase. In addition, we demonstrate the unexpected observation and tunable engineering of robust dual-rail superradiant frequency combs in both the microwave and optical domains. We believe these results pave the way for stable atom-based frequency comb sources, superradiance-based waveform synthesis, and coherent microwave–optical quantum interconnects. 

Our platform opens several promising directions for future exploration. The repetition rate and pulse structure are currently limited by the optical pumping rate required to rebuild population inversion. Incorporating a high-cooperativity optical cavity would Purcell-enhance the repumping process, enabling higher repetition rates and faster control over ensemble participation, while also allowing direct generation of a tunable optical superradiant frequency comb. In parallel, recent advances in superconducting microwave technology offer new opportunities to dynamically engineer the cavity-mediated interactions that govern the collective dynamics. Combined control over ensemble coupling, dissipation, and detuning will unlock a powerful route to programmable quantum light sources, enabling on-demand synthesis of tailored emission waveforms and new paradigms in coherent quantum control.

\begin{acknowledgments}
We acknowledge helpful discussions with Kerry Vahala, Martin Fejer, John Bartholomew, Mi Lei, Andrei Ruskuc, Chun-Ju Wu, Chengyi Luo, Zhiquan Yuan, and Sophie Hermans. \textbf{Funding:} This work was primarily supported by Office of Naval Research grant N00014-22-1-2422. A.F. and J.C. were supported by AFOSR under grant no. FA9550-23-1-0625. J.C. acknowledges support from AFOSR (YIP No. FA9550-25-1-0147) and the Terman Faculty Fellowship at Stanford University. We also acknowledge funding from: US Department of Energy, Office of Science, National Quantum Information Science Research Centers, Co-design Center for Quantum Advantage (contract number DE-SC0012704); Gordon and Betty Moore Foundation Experimental Physics Investigators. R. F. acknowledges support from the Quad fellowship. Fabrication was performed in the Kavli Nanoscience Institute at Caltech. \textbf{Author Contributions:} T.X., J.C. and A.F. conceived the experiments. T.X. and R.F. fabricated the devices. T.X., R.F., and J.L. built the experimental setup. T.X. performed the measurements and analyzed the data. T.X. and W.K.M. performed simulations, and T.X., R.F., W.K.M., J.C., and A.F. interpreted the results. T.X., J.C., and A.F. wrote the manuscript with input from all authors. J.C. and A.F. supervised the project. \textbf{Competing interests:} The authors declare no competing interests. \textbf{Data and materials availability:} The data that support the findings of this study are available from the corresponding author upon reasonable request.

\textbf{Competing interests:} The authors declare no competing interests.

\end{acknowledgments}

\bibliographystyle{naturemag}
\bibliography{ref}

\clearpage

\onecolumngrid
\begin{center}
    {\large Extended Data Figures\par}
\end{center}

\setcounter{figure}{0}

\begin{figure}[h]
\renewcommand{\figurename}{Extended Data Fig.}
\centering
\includegraphics[width=1\linewidth]{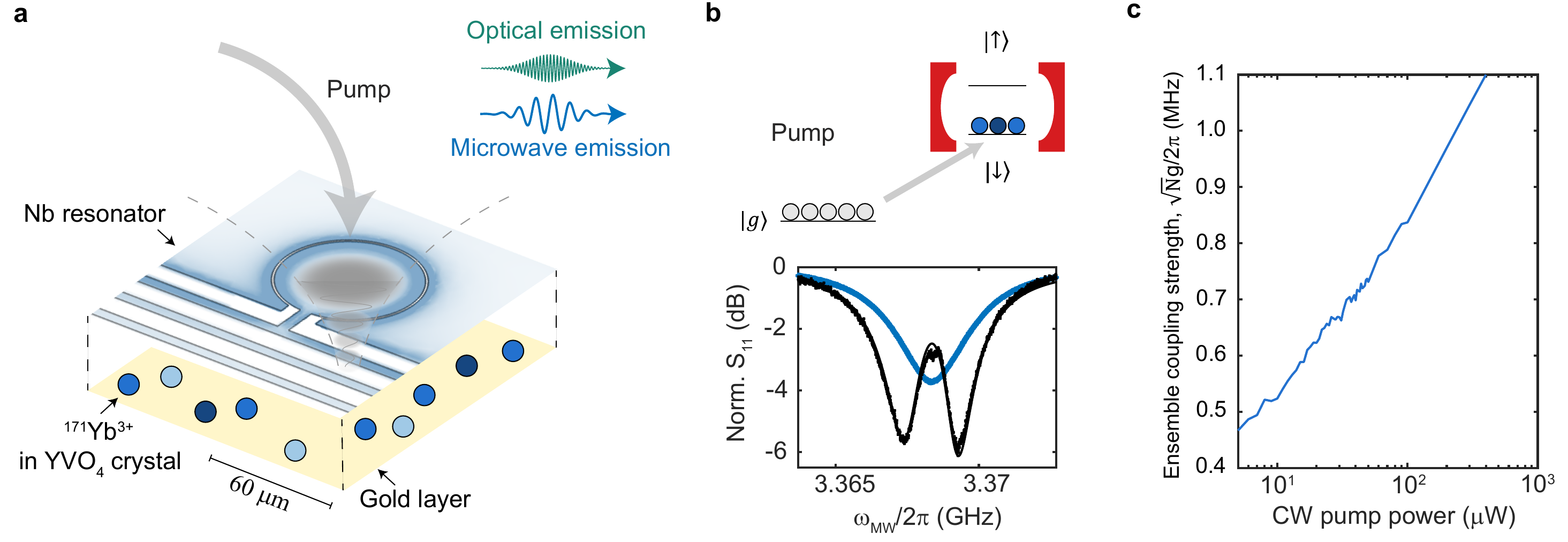}
\caption{
\textbf{Collective coupling between a $^{171}$Yb$^{3+}$ excited-state spin ensemble and a microwave cavity.} \textbf{a}, Detailed device schematic. A niobium film is patterned on a 500 $\mu$m thick $^{171}$Yb$^{3+}$:YVO$_4$ crystal. A 150 nm gold layer is evaporated on the back side, forming a weak optical Fabry-Perot mode between the front and back chip reflections. \textbf{b}, Calibration of spin-cavity coupling strength. By optically pumping the population into the $\ket{\downarrow}$ state, we probe the resonant spin–cavity coupling spectrum and its strength while avoiding superradiant effects. The blue data in the bottom subpanel correspond to the CW pump being off, while the black data are taken with a 400 $\mu$W CW pump. By fitting to a spin-cavity coupling model, we extract an ensemble coupling strength of $\sqrt{N}g \approx 2\pi\times1.1$ MHz, corresponding to a normalized coupling $g_\text{norm} \approx 0.6<1$. The single-spin coupling, $g = 2\pi\times 10$ Hz, is estimated from Comsol simulations, and $N = 10^{10}$. \textbf{c}, Ensemble coupling strength, $\sqrt{N} g$, as a function of pump power. We observe that the ensemble coupling strength increases with pump power, indicating a larger population in the microwave spin manifold.
}
\end{figure}

\begin{figure}
\renewcommand{\figurename}{Extended Data Fig.}
\centering
\includegraphics[width=1\linewidth]{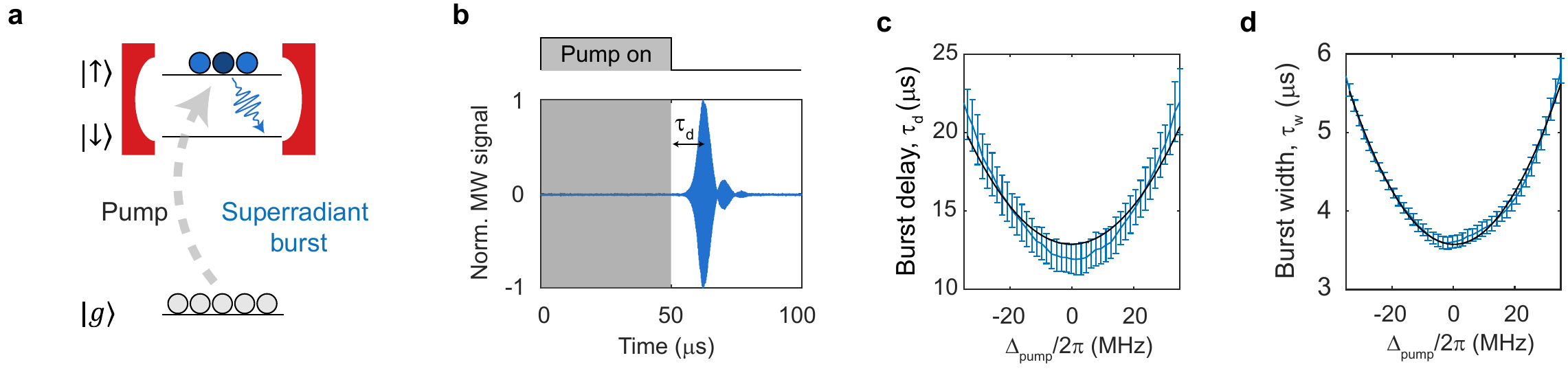}
\caption{
\textbf{Observation and characterization of a delayed superradiant burst under transient excitation.}  
\textbf{a}, A pulsed, transient optical pumping scheme that drives $^{171}$Yb$^{3+}$ spins into the $\ket{\uparrow}$ state. The resulting inverted spin polarization is resonantly coupled to a microwave cavity, leading to a superradiant burst in the microwave domain. 
\textbf{b}, Example of a delayed superradiant burst. A 1 mW pump is applied for 50 $\mu s$, after which a microwave superradiant burst is observed with a delay time $\tau_d$. The superradiant signal is normalized by its peak-to-peak amplitude. 
\textbf{c} and \textbf{d}, Burst delay times, $\tau_d$, and burst width, $\tau_w$, measured as a function of the optical pump detuning, $\Delta_\text{pump}$, which controls the optical transition strength from $\ket{g}$ to $\ket{\uparrow}$. A symmetric dependence of $\tau_d$ and $\tau_w$ is observed, with shorter delay times and narrower widths near resonance. This behavior is consistent with the expected scaling $\tau_d, \tau_w \propto 1/N$, as the pump detuning effectively controls the ensemble size, $N$, in the microwave spin manifold. Data points represent mean values, with error bars indicating the standard deviation over 300 repeated measurements.
}
\end{figure}

\begin{figure}
\renewcommand{\figurename}{Extended Data Fig.}
\centering
\includegraphics[width=1\linewidth]{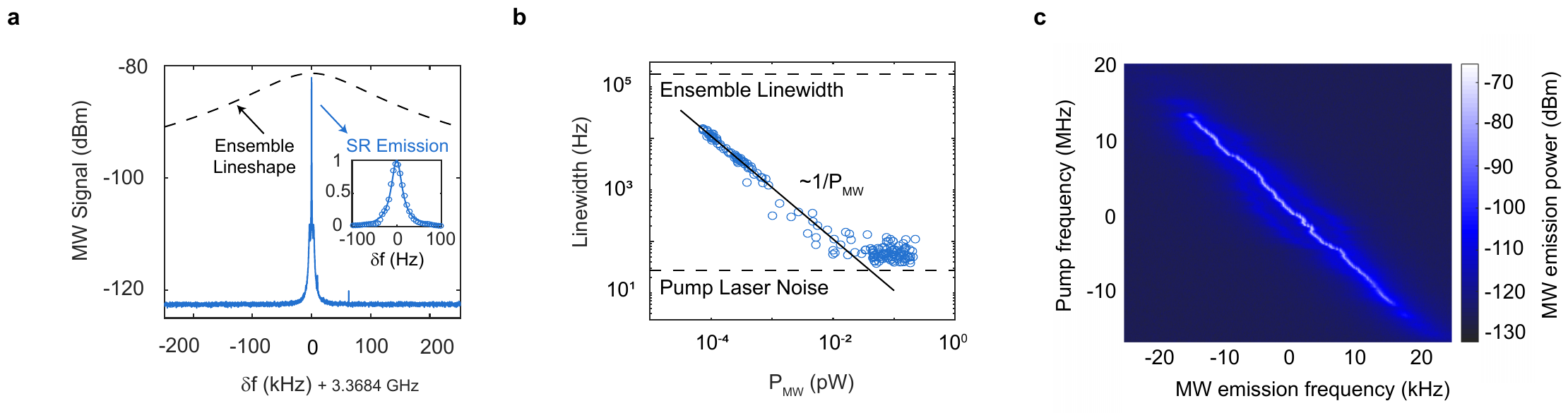}
\caption{
\textbf{Continuous-wave microwave superradiant emission.} 
\textbf{a}, Frequency-domain measurement of CW microwave superradiance, showing a much narrow linewidth of $\approx 30$~kHz. The inhomogeneously broadened ensemble line profile (dashed line) is overlaid for comparison. 
\textbf{b}, Dependence of the linewidth of CW superradiance on emission power, $P_\text{MW}$. An inverse relationship between the linewidth and the emission power is observed, consistent with a noise-broadened Schawlow-Townes linewidth. The lower dashed line denotes the pump laser noise-limited bound. These results confirm that the disordered spin ensemble emits CW superradiance with a much narrower spectral linewidth. 
\textbf{c}, CW superradiance emission frequency at different optical pump detunings. A linear correlation between the microwave emission frequency and optical pump detuning is observed, with a slope of $-1.23$ kHz/MHz.
}
\end{figure}

\begin{figure}
\renewcommand{\figurename}{Extended Data Fig.}
\centering
\includegraphics[width=1\linewidth]{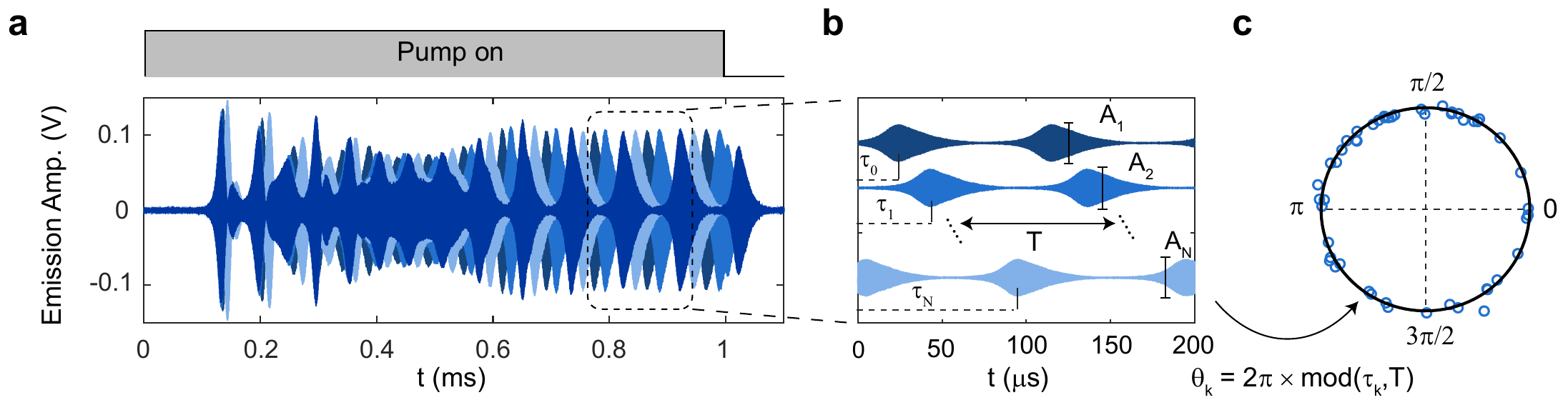}
\caption{
\textbf{Emergence of periodic pulsed superradiance from spontaneous time-translation symmetry breaking.} 
\textbf{a}, Four representative data traces under a 1~ms continuous optical pump. Random onset times of the emergent periodic pulsed phase are observed. 
\textbf{b}, Zoom-in of individual burst trains showing the distribution of burst onset times, $\{ \tau_k \}$, where $k$ denotes the experimental repetition index. After an initial settling time, both the amplitudes, $\{ A_k \}$, and the periodicity, $T$, become independent of $k$, indicating the establishment of a robust pulsed superradiance phase.  
\textbf{c}, Burst onset phases, defined as $\theta_k = 2\pi \times \tau_k/T \; (\text{mod} \; 2\pi)$, measured over 50 repetitions under identical experimental conditions. The phases are uniformly distributed over $2\pi$, demonstrating unbiased spontaneous breaking of continuous time-translation symmetry in the absence of any external timing reference.
}
\end{figure}

\begin{figure}
\renewcommand{\figurename}{Extended Data Fig.}
\centering
\includegraphics[width=0.4\linewidth]{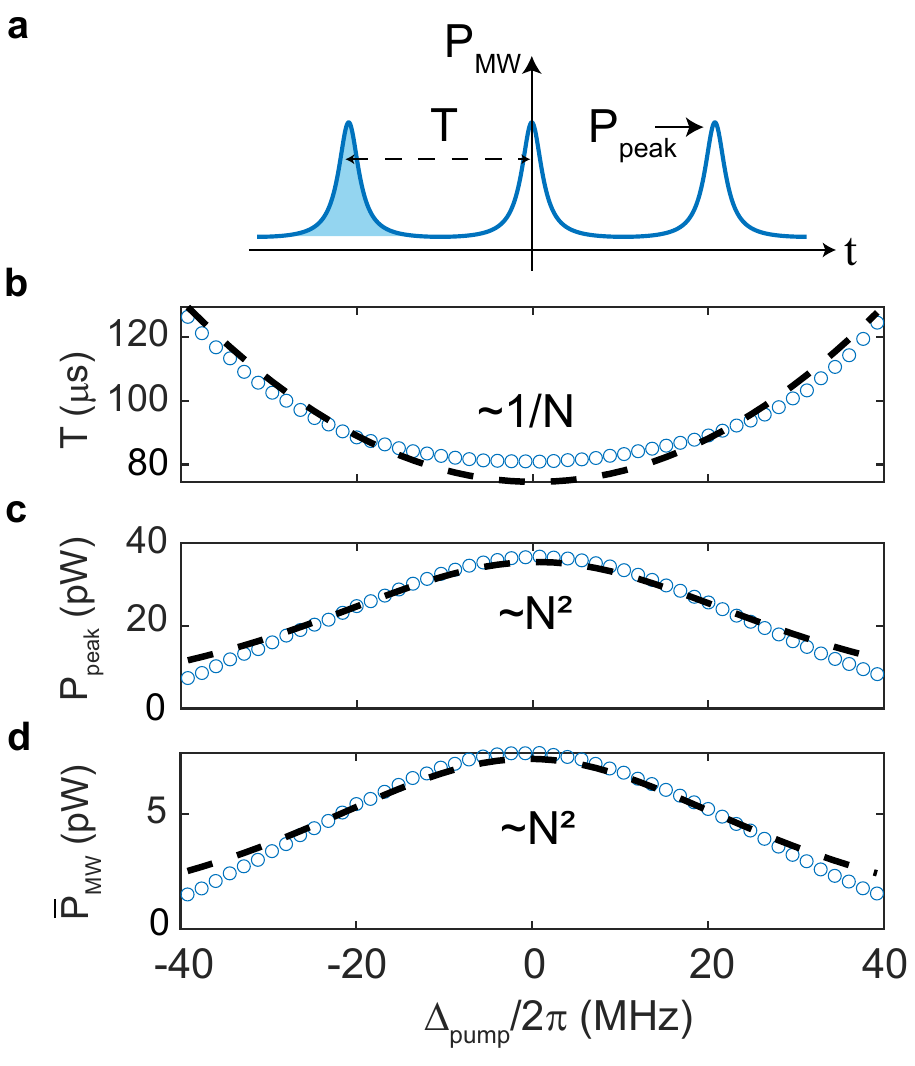}
\caption{\textbf{Scaling behavior of periodic pulsed superradiance.} 
\textbf{a}, Schematic of periodic pulsed superradiance, characterized by three parameters: the pulse-train periodicity $T$, the peak emission power $P_\text{peak}$, and the average emission power $\bar{P}_\text{MW}$. 
\textbf{b-d}, show fits to the experimental data presented in Fig.~3 of the main text, where $T$, $P_{\text{peak}}$, and $\bar{P}_{\text{MW}}$ exhibit good agreement with the expected scaling behaviors of $T \propto 1/N$, $P_{\text{peak}} \propto N^2$, and $\bar{P}_{\text{MW}} \propto N^2$, respectively. Here, $N$ is the effective number of spins in the microwave manifold, tuned via the optical pump detuning, $\Delta_\text{pump}$. When sweeping $\Delta_\text{pump}$, the optical pump power is fixed, corresponding to an excitation over an optical inhomogeneity $\Gamma_\text{op}= 2\pi \times92$ MHz in all three cases. 
}\label{exteded_scaling}
\end{figure}

\begin{figure}
\renewcommand{\figurename}{Extended Data Fig.}
\centering
\includegraphics[width=1\linewidth]{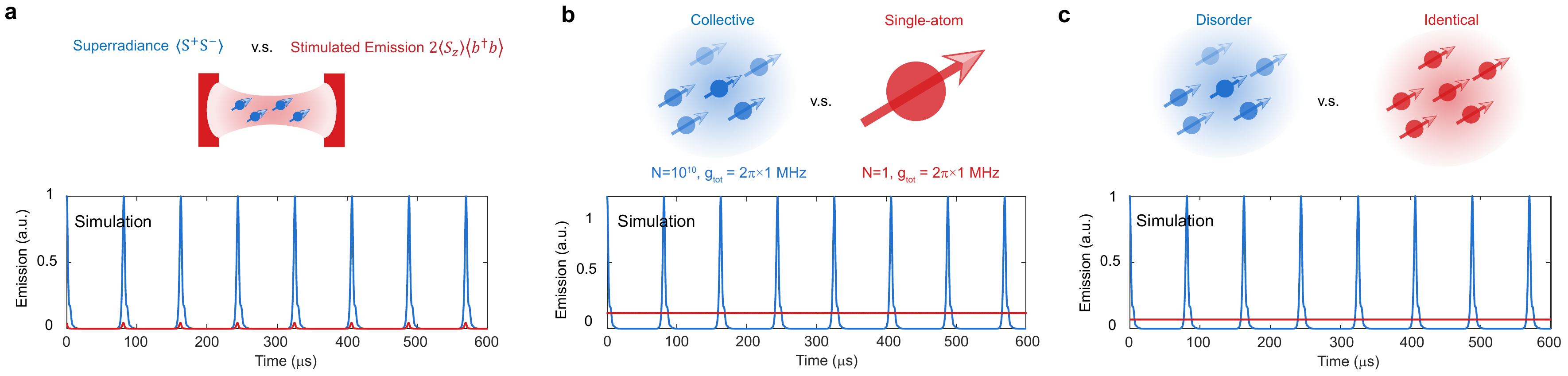}
\caption{
\textbf{Numerical simulation of periodic superradiance using a second-order cumulant expansion method.} 
\textbf{a}, Relative contributions of superradiance (blue) and stimulated emission (red) to the emission strength in the periodic superradiance phase. 
\textbf{b}, Comparison of emission patterns between the many-body case with $N = 10^{10}$ atoms (blue) and the single-atom case with $N = 1$ (red). For a fair comparison, the total effective cavity coupling strength is fixed to $g_{\mathrm{tot}} = \sqrt{N}g = 2\pi \times 1~\text{MHz}$ for both cases. 
\textbf{c}, Comparison of emission patterns in the presence (blue) and absence (red) of spin frequency inhomogeneity, i.e., disorder. All simulations are performed using experimental parameters corresponding to the periodic superradiance phase in Regime III.
}
\end{figure}

\begin{figure}
\renewcommand{\figurename}{Extended Data Fig.}
\centering
\includegraphics[width=1\linewidth]{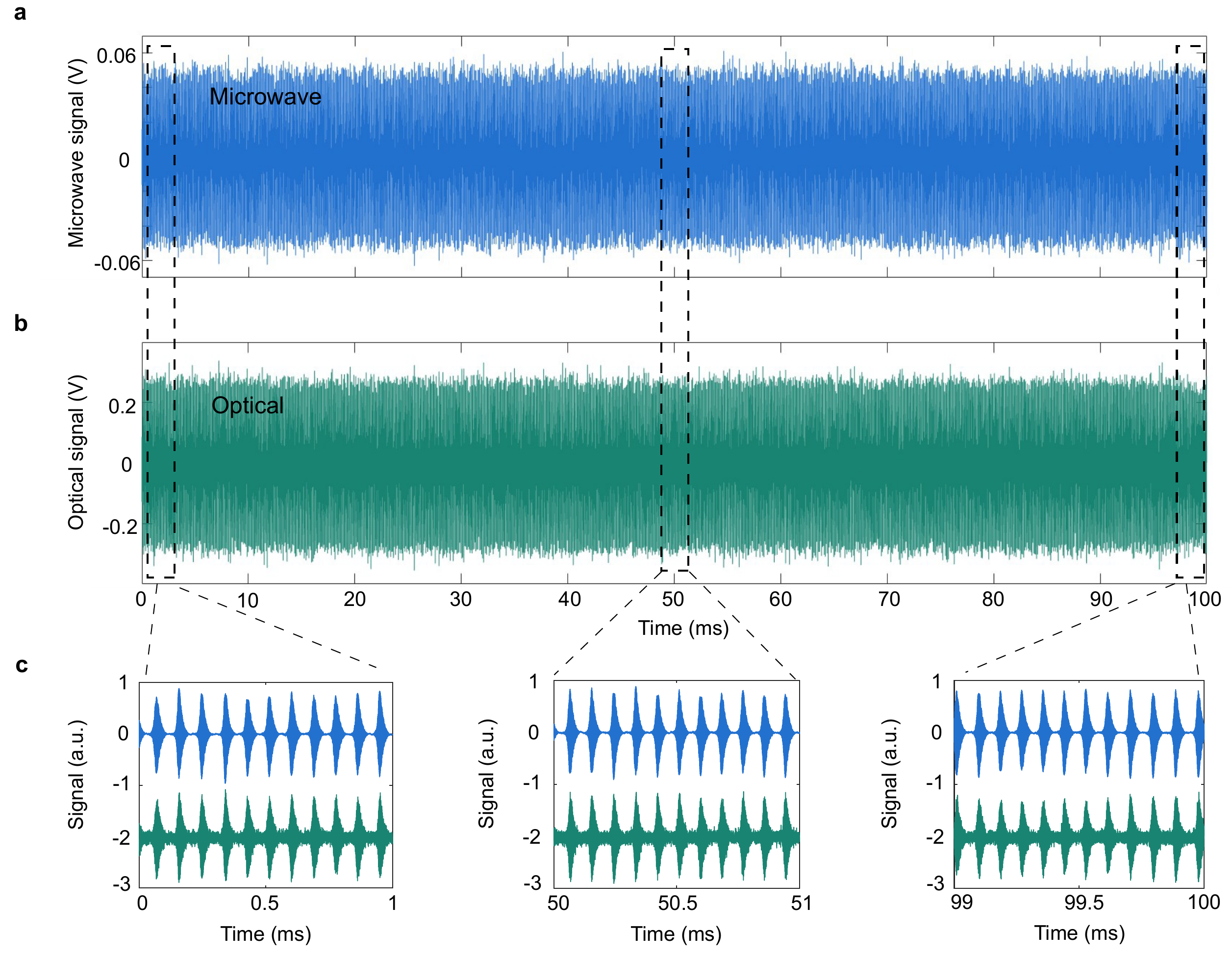}
\caption{
\textbf{Stable, dual-rail periodic pulsed superradiance in both microwave and optical domains.} 
\textbf{a} and \textbf{b}, show simultaneously recorded microwave periodic superradiance and transferred optical periodic superradiance, respectively, over a duration of up to 100~ms. 
\textbf{c}, Zoom-in windows near 1, 50, and 100 ms showing phase- and time-synchronized dual-rail emission patterns, maintaining persistent and robust pulse periodicity without decay. 
}
\end{figure}

\begin{figure}
\renewcommand{\figurename}{Extended Data Table}
\centering
\includegraphics[width=0.7\linewidth]{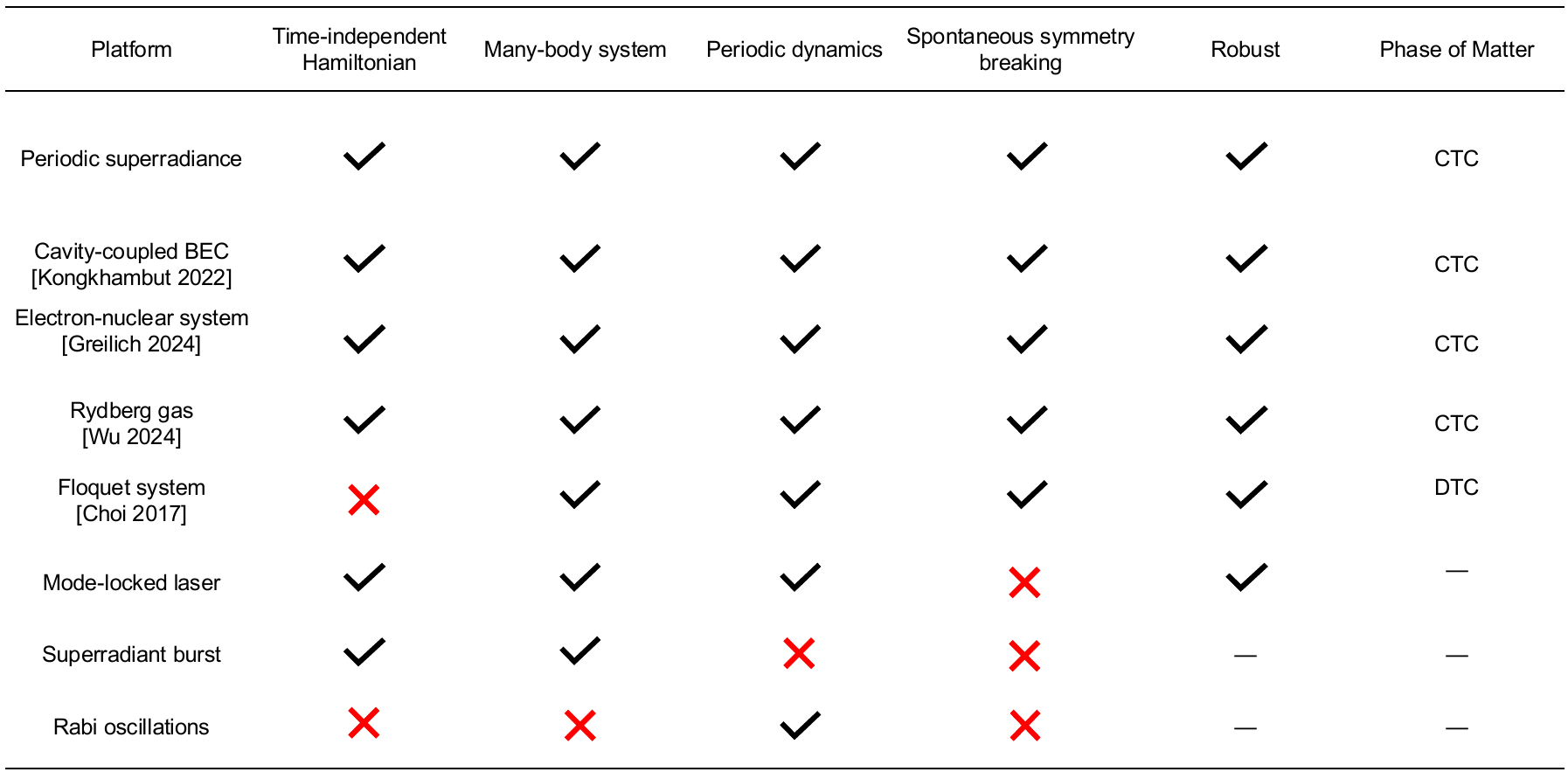}
\caption{
\textbf{Classification of phases of matter in driven quantum and classical systems.} We compare experimental platforms that exhibit periodic pulsed behavior using five criteria: (1) a time-independent Hamiltonian, (2) intrinsically many-body dynamics, (3) periodic temporal behavior, (4) spontaneous symmetry breaking, and (5) robustness to perturbations such as noise and errors. When all five criteria are satisfied simultaneously, the resulting phase can be interpreted as a continuous time crystal (CTC). In contrast, if the Hamiltonian explicitly breaks continuous time-translation symmetry by imposing a time-periodic drive while criteria (2)–(5) remain satisfied, and the system further spontaneously breaks the resulting discrete time-translation symmetry to a subharmonic response, the phase can be identified as a discrete time crystal (DTC). If any of criteria (2)–(5) are not satisfied, the system cannot be regarded as realizing a genuine phase of matter in a driven setting. Some representative references are provided in the table, including Kongkhambut 2022 \cite{kongkhambut2022observation}, Greilich 2024 \cite{greilich2024robust}, Wu 2024 \cite{wu2024dissipative} and Choi 2017 \cite{choi2017observation}.
}
\label{CTC_table}
\end{figure}

\begin{figure}
\renewcommand{\figurename}{Extended Data Table}
\centering
\includegraphics[width=0.7\linewidth]{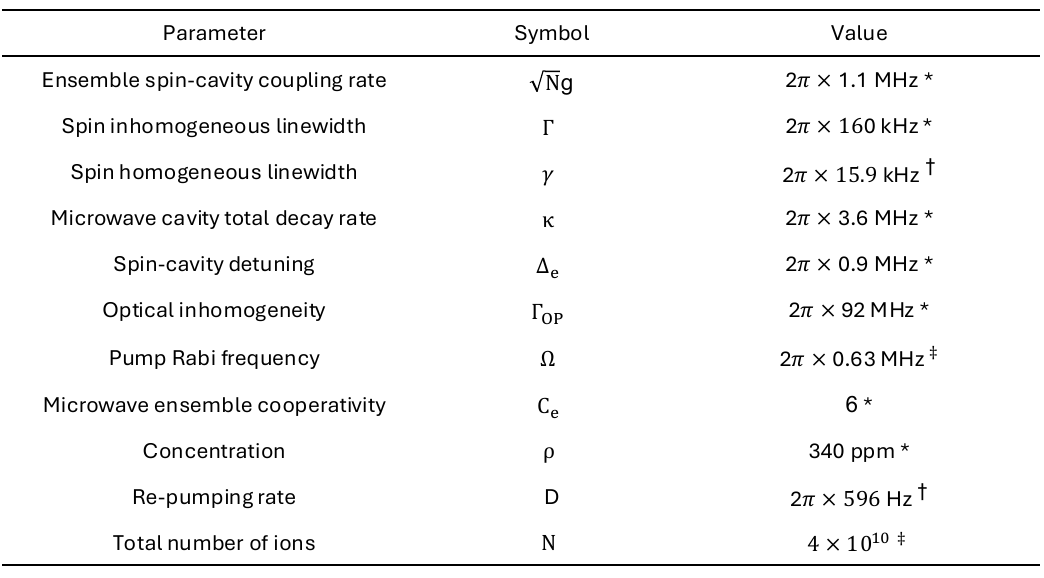}
\caption{
\textbf{Summary of the experimental parameters.} The parameters were determined from experiments ($*$), estimates ($\dagger$), and numerical calculations ($\ddagger $). The ensemble spin-cavity coupling rate, pump Rabi frequency, and microwave ensemble cooperativity are measured under a CW pump power of 400 $\mu$W. 
}
\label{exp_params}
\end{figure}

\clearpage

\renewcommand{\thefigure}{S\arabic{figure}}
\renewcommand{\thetable}{S\arabic{table}}
\renewcommand{\theequation}{S\arabic{equation}}

\begin{center}
    {\large Supplementary Information\par}
\end{center}
\setcounter{figure}{0}

\section{Device fabrication}
$^{171}$Yb$^{3+}$:YVO$_{4}$ crystals are grown by Gamdan Optics and further diced into 6 mm $\times$ 6 mm $\times$ 0.5 mm chips. The ytterbium-171 concentration is measured to be 340 ppm via secondary ion mass spectrometry (SIMS). A 150 nm niobium film is sputtered onto the chip surface at 3.5 mTorr. The device patterns are written by an electron beam lithography system (Raith EBPG 5200) with ma-N 2403 photoresist. A 25 nm aluminum layer is evaporated as a hard mask and further lift-off under remover PG. The niobium is dry-etched in plasma reactive-ion etcher and the aluminum mask is removed using a wet-etched process with aluminum etchant and 3.5$\%$ hydrofluoric acid. Finally, a 100 nm gold layer is evaporated on the back of the chip to form a reflective coating.   

\section{Experimental setup}
A master laser (Toptica DL Pro) is locked to a stable cavity reference (Stable laser systems) via the Pound-Drever-Hall (PDH) method. The experiment laser (Moglabs CEL) is offset-locked to the master laser using a Vescent D2-135 servo and TPI 1005-A frequency source for further tuning the locking frequency. The chip is mounted on the mixing-chamber plate of a dilution refrigerator (Bluefors LD250). A total of 70-dB attenuation is distributed along the microwave input lines. The microwave output line is connected to cryogenic circulators (LNF-CIC2.8-4.3A) at the mixing chamber plate and linked to a high-electron-mobility transistor (HEMT) amplifier (LNF-LNC2-6B) at 4K plate via superconducting coax cables. The microwave signal outside the fridge is measured on a Field Fox N9115A spectrum analyzer and an oscilloscope (Tektronics TDS7104). For pulsed measurements, a digitizer (Alazar ATS9130) is used to acquire data after mixing down to 5 MHz with an IQ mixer (Marki Microwave IQ0307 LXP). 

The optical pump is delivered into the device through a pair of aspheric lenses and collected along the same path via the gold layer reflection. The beam is focused on the gold layer with a 4 $\mu$m beam waist, resulting in a 20 $\mu$m beam waist on the front surface of the chip for the experiment. The collected optical photons are combined with another optical path outside the fridge for heterodyne measurements at a 200 MHz offset frequency. The optical heterodyne signal is detected by a high-speed photodiode (ALPHALAS UPD-35-IR-2) and further analyzed using Field Fox N9115A spectrum analyzer and Tektronix TDS7104 oscilloscope. For pulsed excitations, an acousto-optic modulator (AOM, Aerodiode model 6) shapes the optical pump beam with a trigger from Zurich HDAWG. All electronics are externally locked to a 10 MHz rubidium frequency standard (SRS FS725). All the relevant experimental parameters can be found in Extended Data Table~\ref{exp_params}.

\section{System modeling}
\subsection{Mean-field equations.}
Our cavity-coupled, coherent spin dynamics, associated with the dynamics in the excited-state microwave transition manifold, are modeled as an effective two-level system, \{$\ket{\downarrow},\ket{\uparrow}$\}, with the following Hamiltonian:
\begin{equation}
\begin{split}
\hat{\mathcal{H}}& = \hbar \omega_e \hat{b}^\dagger \hat{b} + \sum_k E_{\downarrow,k}\hat{\sigma}_{\downarrow\downarrow,k} + \sum_k E_{\uparrow,k}\hat{\sigma}_{\uparrow\uparrow,k}\\
& +\sum_k \hbar g_k\parens{\hat{b}^\dagger \hat{\sigma}_{\downarrow\uparrow,k}+\hat{\sigma}_{\uparrow\downarrow,k}\hat{b}} \\
\end{split}
\end{equation}
where $\omega_e$ is the microwave resonator frequency, $\hat{b}$ and $\hat{b}^{\dagger}$ are the resonator annihilation and creation operators, $E_{\downarrow,k}$ and $E_{\uparrow,k}$ are the energies of the $\ket{\downarrow}$ and $\ket{\uparrow}$ states of the $k$-th spin, $g_k$ is the single-ion coupling strength between the cavity and the $k$-th spin, and $\hat{\sigma}_{\mu\nu, k} = \ket{\mu}_k\bra{\nu}$ with $\mu, \nu = \downarrow, \uparrow$. 
To gain an intuitive understanding, we first apply a mean-field approximation to the Heisenberg equation of motion for the spin ($\hat{\sigma}$) and cavity ($\hat{b}$) operators, which are given by:
\begin{equation}
\begin{split}
    \langle \dot{\hat{b}}\rangle &= -i\omega_e \langle \hat{b}\rangle -\frac{\kappa}{2}\langle \hat{b}\rangle-i\sum_k g_k \langle \hat{\sigma}_{\downarrow\uparrow,k}\rangle + F_e(t) \\ 
    \langle \dot{\hat{\sigma}}_{\downarrow\uparrow,k} \rangle &= -i\omega_{k}\langle \hat{\sigma}_{\downarrow\uparrow,k}\rangle -\parens{\frac{\gamma}{2}+\frac{\gamma_1+D}{2}} \langle \hat{\sigma}_{\downarrow\uparrow,k}\rangle\\
    &+ ig_k \parens{\langle\hat{\sigma}_{\uparrow\uparrow,k}\rangle-\langle\hat{\sigma}_{\downarrow\downarrow,k}\rangle}\langle\hat{b}\rangle\\    
    \langle\dot{\hat{\sigma}}_{\uparrow\uparrow,k}\rangle &= D\langle\hat{\sigma}_{\downarrow\downarrow,k}\rangle - \gamma_1 \langle\hat{\sigma}_{\uparrow\uparrow,k}\rangle\\
    &- ig_k \parens{\langle\hat{\sigma}_{\uparrow\downarrow,k}\rangle\langle\hat{b}\rangle-\langle\hat{b}^{\dagger}\rangle\langle\hat{\sigma}_{\downarrow\uparrow,k}\rangle}\\
    \langle \dot{\hat{\sigma}}_{\downarrow\downarrow,k}\rangle &= -D\langle \hat{\sigma}_{\downarrow\downarrow,k}\rangle + \gamma_1 \langle \hat{\sigma}_{\uparrow\uparrow,k}\rangle\\
    &+ ig_k \parens{\langle \hat{\sigma}_{\uparrow\downarrow,k}\rangle\langle \hat{b}\rangle-\langle \hat{b}^{\dagger}\rangle\langle \hat{\sigma}_{\downarrow\uparrow,k}\rangle}\\
\end{split}
\label{meanfield-eqn}
\end{equation}
Here, $\kappa$ is the cavity decay rate, $F_e(t)$ represents time-dependent random noise fluctuations in the cavity field, which induce random kicks that drive the population-inverted Bloch spin vectors away from $z$-axis, $\omega_{k}$ is the $k$-th spin transition frequency, $\gamma$ is the homogeneous spin dephasing rate, $D$ is the incoherent pumping rate into the $\ket{\uparrow}$ state, and $\gamma_1$ is the spin relaxation rate. Note that the equations of motion include incoherent processes such as pumping, cavity dissipation, and noise terms, thereby capturing the intrinsic openness of cavity QED systems. In particular, the spin inversion, or equivalently the spin polarization toward the $\ket{\uparrow}$ state, is determined by the balance between incoherent pumping and the relaxation process.

In the bad cavity regime, where the cavity decay rate, $\kappa$, is much greater than the collective coupling strength, $\sqrt{N}g$, we can adiabatically eliminate the cavity field, which yields a reduced set of equations of motion involving only the spin degrees of freedom:
\begin{equation}
\begin{split}
    \langle \dot{\hat{\sigma}}_{\downarrow\uparrow,k}\rangle &= -i\delta_{k}\langle \hat{\sigma}_{\downarrow\uparrow,k}\rangle - \parens{\frac{\gamma}{2}+\frac{\gamma_1+D}{2}} \langle \hat{\sigma}_{\downarrow\uparrow,k}\rangle\\
    &- \frac{g^2}{i\Delta_e+\frac{\kappa}{2}}\langle \hat{\sigma}_{z,k}\rangle\langle \hat{S}_{_{\downarrow\uparrow}}\rangle + F_k(t)
    \\
    \langle \dot{\hat{S}}_{z}\rangle &=  -\parens{D-\gamma_1}N - \parens{D+\gamma_1}\langle \hat{S}_z\rangle \\
    &+ \frac{2g^2\kappa}{\Delta_e^2+(\frac{\kappa}{2})^2}\langle \hat{S}_{\uparrow\downarrow}\rangle\langle \hat{S}_{\downarrow\uparrow}\rangle + F_z(t)
\end{split}
\end{equation}
Here, $\hat{S}_z = \frac{1}{2}\sum_k \hat{\sigma}_{z,k}$ is the collective spin polarization along the $z$-axis with $\hat{\sigma}_{z,k} = \hat{\sigma}_{\downarrow\downarrow,k}-\hat{\sigma}_{\uparrow\uparrow,k}$, $F_k(t)=\left(\frac{-ig\langle\hat{\sigma}_{z,k}\rangle}{i\Delta_e+\kappa/2}\right)F_e(t)$ and $F_z(t) = 2\text{Re}\left[\left(\frac{ig\langle\hat{S}_{\uparrow\downarrow}\rangle}{i\Delta_e+\kappa/2}\right)F_e(t)\right]$ are the spin noise fluctuations arising from $F_e(t)$, and $\delta_k=\omega_k-\omega_0$ is the detuning of the $k$-th spin relative to the center of the inhomogeneous ensemble line profile. Since, in our experiment, the microwave cavity field is approximately homogeneous inside the beam waist~\cite{xie2025scalable}, we assume that the cavity coupling strength is homogeneous across all spins, i.e., $g_k = g$.
Note that the rate of change of an individual spin’s coherence, $\langle \dot{\hat{\sigma}}_{\downarrow\uparrow,k} \rangle$, depends on the collective ensemble coherence, $\langle \hat{S}_{\uparrow\downarrow} \rangle$, as a result of cavity-mediated coupling. In the presence of ensemble coherence, spin inversion ($\langle \hat{\sigma}_{z,k} \rangle < 0$), acting in tandem, leads to the growth of individual spin coherence. At the collective level, the ensemble polarization dynamics, $\langle \dot{\hat{S}}_z \rangle$, depend on $\langle \hat{S}_{\uparrow\downarrow} \rangle \langle \hat{S}_{\downarrow\uparrow} \rangle$, which is responsible for superradiant decay when this term is dominant.

\subsection{Second-order cumulant expansion.}
While the mean-field equations provide a leading-order, intuitive understanding of superradiant cavity-QED dynamics with a disordered spin ensemble, the numerically simulated periodicity of pulsed superradiance is inconsistent with the experimental observations (Supplementary Information). 

To address this discrepancy, we employ a second-order cumulant expansion~\citeSI{kubo1962generalized} in numerical simulations to more accurately capture the spin–spin correlations responsible for the superradiance process. Since random on-site disorder breaks the permutation symmetry among the spins, it prevents us from utilizing collective spin operators for the numerical simulation, thereby requiring full many-body simulations~\citeSI{debnath2019collective}. Specifically, to ensure that the second-order expansion forms a closed set of equations, we approximate expectation values of third-order or higher-order operators as combinations of lower-order expectation values~\citeSI{plankensteiner2022quantumcumulants}:
\begin{equation}
\begin{split}
\langle \hat{A}_1 \hat{A}_2 \hat{A}_3 \rangle &= \langle \hat{A}_1\rangle\langle \hat{A}_2 \hat{A}_3\rangle + \langle \hat{A}_2\rangle \langle \hat{A}_1 \hat{A}_3 \rangle \\
&+ \langle \hat{A}_3\rangle\langle \hat{A}_1 \hat{A}_2\rangle -2\langle \hat{A}_1\rangle\langle \hat{A}_2\rangle\langle\hat{A}_3 \rangle
\end{split}
\end{equation}
Based on this approximation, together with the assumption that phase averages vanish, all relevant third-order expectation values are expressed in terms of second-order or local expectation values:
\begin{equation}
\begin{split}
\langle \hat{\sigma}_{z,k}\hat{b}^\dagger \hat{b} \rangle &= \langle \hat{\sigma}_{z,k} \rangle\langle \hat{b}^\dagger \hat{b}\rangle\\
\langle \hat{\sigma}_{z,k}\hat{b}^\dagger \hat{\sigma}_{\downarrow\uparrow,j} \rangle &= \langle \hat{\sigma}_{z,k} \rangle\langle \hat{b}^\dagger \hat{\sigma}_{\downarrow\uparrow,j}\rangle\\
\langle \hat{\sigma}_{z,k} \hat{\sigma}_{\uparrow\downarrow,j} \hat{b}\rangle &= \langle \hat{\sigma}_{z,k} \rangle\langle \hat{\sigma}_{\uparrow\downarrow,j}\hat{b}\rangle \\
\end{split}
\end{equation}
Based on this second-order expansion, each spin evolves according to the equations of motion below:  
\begin{equation}
\begin{split}
    \frac{d}{dt}&\langle \hat{b}^\dagger\hat{\sigma}_{\downarrow\uparrow,k}\rangle = \parens{i\omega_e-i\omega_{k}-\frac{\kappa+\gamma+\gamma_1+D}{2}} \langle \hat{b}^\dagger\hat{\sigma}_{\downarrow\uparrow,k}\rangle\\
    &+ ig \parens{\frac{1-\langle \hat{\sigma}_{z,k}\rangle}{2}+\sum_{j\neq k}\langle \hat{\sigma}_{\uparrow\downarrow,j}\hat{\sigma}_{\downarrow\uparrow,k}\rangle-\langle \hat{\sigma}_{z,k}\rangle\langle \hat{b}^\dagger \hat{b}\rangle} \\
    \frac{d}{dt}&\langle \hat{b}^\dagger \hat{b}\rangle = -\kappa \langle \hat{b}^\dagger \hat{b}\rangle+\kappa N_{th}\\
    &+ig\parens{\sum_k\langle \hat{\sigma}_{\uparrow\downarrow,k}\hat{b}\rangle-\sum_k \langle \hat{b}^\dagger \hat{\sigma}_{\downarrow\uparrow,k}\rangle} \\
    \frac{d}{dt}&\langle \hat{\sigma}_{\uparrow\downarrow,i}\hat{\sigma}_{\downarrow\uparrow,j}\rangle = \parens{i\omega_i-i\omega_j-\gamma-\gamma_1-D}\langle \hat{\sigma}_{\uparrow\downarrow,i}\hat{\sigma}_{\downarrow\uparrow,j}\rangle \\
    &+ig\langle \hat{\sigma}_{z,i}\rangle\langle \hat{b}^\dagger \hat{\sigma}_{\downarrow\uparrow,j}\rangle-ig\langle \hat{\sigma}_{z,j}\rangle\langle \hat{\sigma}_{\uparrow\downarrow,i}\hat{b}\rangle\\
    \frac{d}{dt}&\langle \hat{\sigma}_{z,k}\rangle = -D\parens{\langle \hat{\sigma}_{z,k}\rangle+1}+\gamma_1\parens{1-\langle \hat{\sigma}_{z,k}\rangle}\\
    &-2ig\parens{\langle \hat{b}^\dagger \hat{\sigma}_{\downarrow\uparrow,k}\rangle-\langle \hat{\sigma}_{\uparrow\downarrow,k}\hat{b}\rangle}\\
\end{split}
\label{2nd-simu-method}
\end{equation}
where $N_\text{th}$ denotes the thermal photon noise, which sets the noise level of the microwave cavity photon number. Note that, under the bad-cavity assumption, the equation for $\langle\hat{b}^\dagger \hat{b}\rangle$ simplifies to:
\begin{equation}
\begin{split}
    \frac{d}{dt}\langle\hat{b}^\dagger \hat{b}\rangle &= -\kappa \langle\hat{b}^\dagger \hat{b}\rangle + \kappa N_{th}\\
    &+ \frac{4g^2}{\kappa_{tot}}
    \parens{\frac{N}{2}- \langle \hat{S}_z\rangle
     - 2\langle \hat{S}_z \rangle  \langle \hat{b}^\dagger \hat{b} \rangle
     + \langle \hat{S}_{\uparrow\downarrow}\hat{S}_{\downarrow\uparrow}\rangle}
\end{split}
\end{equation}
where $\kappa_\text{tot} = \kappa\frac{1}{1+(2\Delta_e/\kappa)^2}$ denotes the effective total decay rate, where $\Delta_e = \omega_s - \omega_e$ is the detuning of the central frequency of the inhomogeneous broadening relative to the cavity resonance frequency. In the total decay rate expression, we neglect the detuning dependence for individual spins, as the bad-cavity dissipation rate ($\kappa = 2\pi \times 3.6$ MHz) dominates over the inhomogeneous ensemble linewidth ($\Gamma = 2\pi \times 160$ kHz). In the main text, we use $\kappa$ to denote $\kappa_{\text{tot}}$ for simplicity. Importantly, $\langle \hat{S}_{\uparrow\downarrow}\hat{S}_{\downarrow\uparrow}\rangle = \sum_k\sum_{j\neq k}\langle \hat{\sigma}_{\uparrow\downarrow,j}\hat{\sigma}_{\downarrow\uparrow,k}\rangle$ represents the collective spin-spin correlation that scales with $N^2$. 

\section{Derivation of the periodicity}
As discussed in the main text, the time-periodic superradiant pulse emission arises from persistent spin dynamics in which coherence growth, precession, and dephasing recur cyclically, while optical repumping between cycles replenishes the spin population and enables the buildup required for the next superradiant burst. The periodicity of a superradiant burst train, $T$, can be analytically calculated by treating the repumping process classically, yielding:
\begin{equation}
    T = \frac{1}{D} \log(\frac{N_f-N_i}{N_f-\frac{\kappa \Gamma }{4g ^2}}),
    \label{eq:burst_period}
\end{equation}
where $D$ is the repumping rate, $N_i$ is the initial population of the spin manifold, and $N_f$ is the final saturated population reached in the absence of superradiant bursts. The emission threshold occurs when the ensemble cooperativity reaches unity, where coherence growth balances the spin dephasing rate. Since the ensemble cooperativity, $C = \frac{4Ng^2}{\kappa \Gamma}$, scales with $N$ and assuming $N_i \ll N_f$, Eq.~(\ref{eq:burst_period}) can be approximately simplified as
\begin{align}
    T \approx \frac{1}{D}\log(\frac{C_f}{C_f-1}) \approx \frac{1}{DC_f},
\end{align}
where the ensemble cooperativity $C_f$ corresponds to the final cooperativity in the absence of superradiant bursts, which can be calibrated by pumping the population into the $\ket{\downarrow}$ state (Extended Data Fig.~1). This reveals a cooperativity-dominated burst periodicity, $T$, leading to the universality observed in the pulsed superradiance phase (Regime III), as presented in Fig.~3 of the main text.

\section{Delayed superradiant bursts}
% Here, we discuss the physics of a delayed superradiant burst arising from transient pulsed excitation, serving as an experimental smoking gun for the predominance of superradiance (see Ref.~\citeSI{angerer2018superradiant} for theoretical details). For the case of a spin ensemble subject to an initial transient pump, one can use the same equations of motion presented earlier by simply setting the repumping rate, $D = 0$. We also ignore the dephasing ($\gamma$) and cavity noise ($\eta_b$) terms, as they are the least dominant contributions. 
For the case of a spin ensemble subject to an initial transient pump, the collective spin polarization, $\hat{S}_z$, evolves according to the following equation of motion~\citeSI{angerer2018superradiant}:
\begin{equation}
    \langle\dot{\hat{S}}_z(t)\rangle= \frac{4g^2}{\kappa}\left(\frac{N}{2}-\langle\hat{S}_z(t)\rangle\right)\left(\frac{N}{2}+\langle\hat{S}_z(t)\rangle+1\right),
\end{equation}
where a collective spin state with total spin $S = \frac{N}{2}$ and projection $S_z$ is used. An analytical solution can be obtained by solving the above equation assuming $N \gg 1$, which yields
\begin{equation}
    \langle \hat{S}_z (t) \rangle = \frac{N}{2}\tanh\left(\frac{2Ng^2}{\kappa}(t-\tau_d)\right),
\end{equation}
where $\tau_d \approx \frac{\kappa}{2Ng^2}\log(|\tan(\frac{\theta}{2})|) $ is the delay time of a superradiance burst, where $\theta = \cos^{-1}(2\langle\hat{S}_z(0) \rangle/N)$. The duration of the burst, $\tau_w$, is given by $\tau_w = \frac{\kappa}{Ng^2}\log(\sqrt 2 +1)$. Note that both the delay time, $\tau_d$, and the burst width, $\tau_w$, scale inversely proportional to the emitter ensemble size, $N$, i.e., $\tau_d, \tau_w \propto 1/N$. Experimental results showing good agreement with the theoretical predictions are presented in Extended Data Fig. 2.

\section{Scaling of periodic superradiant pulse trains}
% Here, we discuss the scaling behavior of the peak emission strength and the periodicity of superradiant burst trains (Regime III). 
Our theoretical framework predicts how the burst periodicity ($T$), peak power ($P_{\text{peak}}$), and average emission power ($\bar{P}_{\text{MW}}$) scale with the number of participating emitters, $N$ (Extended Data Fig.~\ref{exteded_scaling}a). First, the burst periodicity is governed by the ensemble cooperativity, leading to the scaling of $T \propto 1/N$ (Extended Data Fig.~\ref{exteded_scaling}b). Second, each burst originates from the collective emission of an inverted ensemble of $N$ atoms, resulting in the emission of $N$ photons. Since the superradiant burst duration, $\tau_w$, scales inversely with the ensemble size, i.e., $\tau_w \propto 1/N$, the peak power---proportional to the burst energy divided by its duration---scales quadratically with $N$, i.e., $P_{\text{peak}} \propto N^2$ (Extended Data Fig.~\ref{exteded_scaling}c). Third, the average emission power follows from the burst energy multiplied by the repetition rate, leading to the same quadratic scaling with $N$, i.e., $\bar{P}_{\text{MW}} \propto N^2$ (Extended Data Fig.~\ref{exteded_scaling}d).

As shown in Extended Data Fig.~\ref{exteded_scaling}, all these scaling predictions are in excellent agreement with the experimental data. This agreement enables a universal rescaling of the experimentally observed superradiant pulse trains in terms of the effective ensemble size $N$, or equivalently, the experimentally tunable cooperativity.

\section{Linewidth of superradiance emission}
The emission linewidth of a superradiant atomic ensemble has been studied extensively in the literature, both in the thermal ground-state and finite-temperature regimes~\citeSI{jin2015proposal,bohnet2012steady,loughlin2023quantum}. By linearizing the quantum Langevin equations in the frequency domain, one can compute the phase fluctuation correlation function of the emitted field and hence derive the emission linewidth as 
\begin{equation}
    \Delta f = \frac{1}{4\pi n_c \kappa_c} \left(\frac{\kappa_a\kappa_c}{\kappa_a+\kappa_c}\right)^2\left(n_{th}+\frac{1}{2}+n_{sp}+\frac{1}{2}\right)
    \label{ST_linewidth}
\end{equation}
where $n_c$ is the intracavity photon number, $\kappa_c$ and $\kappa_a$ are the linewidths of the cavity mode and atomic ensemble, respectively, $n_{th}$ is the thermal occupancy of the cavity mode, and $n_{sp}$ is the spontaneous emission factor. The linewidth is therefore determined by noise contributions from both the thermal bath of the cavity mode and the imperfect inversion of the atomic ensemble. In the ideal limit of a quantum ground state with perfect population inversion, one recovers the result of Ref.~\citeSI{bohnet2012steady} by setting $n_{th}=n_{sp}=0$. Noting that the emission power $P$ is proportional to the intracavity photon number, i.e., $P \propto n_c$, the linewidth correspondingly narrows as $\Delta f_{ST} \propto 1/P$. 

By fitting to Eq.~(\ref{ST_linewidth}) to the data shown in Extended Data Fig.~3, we extract a total noise contribution of $n_{th}+n_{sp} = 5.8$ quanta. From numerical simulations, the spontaneous emission factor is estimated to be $n_{sp}= 2.6$ quanta, leaving $n_{th} = 3.2$ quanta for the thermal occupancy of the microwave resonator, corresponding to an effective resonator temperature of $\approx$0.6K. 

We also observe a correlation between the microwave and optical transitions frequencies consistent with previous observations~\citeSI{bartholomew2020chip}. The laser linewidth is carefully calibrated using the Pound-Drever-Hall error signal, yielding a locking linewidth of 22 kHz. This corresponds to a 27 Hz limit on the microwave emission frequency, as shown in Extended Data Fig. 3b. 

On the optical side, the transferred superradiant emission inherits the linewidth of the optical pump laser due to energy conservation. The data shown in Fig.~4D of the main text correspond to heterodyne detection between the optical emission and the pump laser, which in principle should yield the same linewidth as the microwave comb ($\approx50$~Hz). However, the measured optical comb linewidth is $\approx1$~kHz, likely due to frequency jitter arising from the time delay between the optical path through the dilution refrigerator and the local oscillator path.

\section{Criteria for a continuous time-crystal phase}
We evaluate whether the observed periodic pulsed superradiance can be classified as a continuous time-crystal phase by applying a set of five criteria commonly associated with time-crystalline order~\citeSI{kongkhambut2022observation,wu2024dissipative,greilich2024robust,tucker2018shattered}.

First, the underlying Hamiltonian should be time-independent, ensuring that any emergent periodic behavior is not trivially imposed by external driving. Second, the observed dynamics should arise from genuine many-body interactions, rather than from single-particle effects or externally enforced synchronization. Third, the system should exhibit persistent and well-defined periodic temporal behavior in its steady state. Fourth, this periodicity should correspond to the spontaneous breaking of continuous time-translation symmetry, meaning that the system evolves with a characteristic frequency that is not explicitly set by the Hamiltonian. Fifth, the observed behavior should be robust against perturbations such as noise, dissipation, and experimental imperfections, indicating the presence of a stable dynamical phase rather than fine-tuned transient dynamics. 

We systematically evaluate these criteria across different experimental platforms and summarize the results in Extended Data Table 1. This comparison provides a unified framework for assessing whether periodic superradiant dynamics can be understood within the broader context of continuous time crystals.

To further corroborate that our system spontaneously breaks continuous time-translational symmetry, we apply a transient optical pump for 1~ms and monitor the emergence of the periodic pulsed-emission phase. The temporal phase, $\theta_k$, of the continuous time crystal is defined from the onset time of the periodic burst oscillations as $\theta_k = 2\pi \times \tau_k/T \; (\text{mod} \; 2\pi)$, where $k$ indexes individual experimental realizations and $\tau_k$ is the burst onset time in the $k$-th run. As shown in Extended Data Fig.~4, the system first undergoes superradiant burst dynamics and then relaxes into a long-lived periodic pulsed phase that persists without decay over the observation window. Crucially, the onset time, and hence the phase $\theta_k$, varies randomly across repeated experiments, indicating that the phase is not determined by the transient pump. This uniform randomness provides evidence that temporal order emerges spontaneously, consistent with the spontaneous breaking of continuous time-translational symmetry.

In addition, as shown in Extended Data Fig. 7, the observed periodic superradiance represents a robust dynamical phase in which both microwave superradiance and the transferred optical superradiance exhibit essentially non-decaying oscillations under continuous pumping.

\section{Cavity QED dynamics modeling and numerical simulation details}
\subsection{Modeling microwave and optical superradiance dynamics via the mean-field approach}
To model the microwave superradiance and transferred optical superradiance together, we need to consider the full three-level system, $ \{\ket{\uparrow}, \ket{\downarrow}, \ket{g} \} $. To this end, we extend the microwave-only Hamiltonian in Methods to incorporate the optical coherent dynamics as follows:
\begin{equation}
\begin{split}
\hat{\mathcal H} & =  \hbar \omega_{o,c} \hat{a}^{\dagger}\hat{a} + \hbar \omega_{e,c} \hat{b}^{\dagger}\hat{b}  + \sum_{k} E_{g,k}\hat{\sigma}_{gg,k} + \sum_{k} E_{\downarrow,k}\hat{\sigma}_{\downarrow\downarrow,k} + \sum_{k} E_{\uparrow,k}\hat{\sigma}_{\uparrow\uparrow,k}
\\
& + \sum_{k} \hbar g_{o,k} \parens{\hat{a}^{\dagger} \hat{\sigma}_{g\downarrow,k} + \hat{a}\hat{\sigma}_{\downarrow g,k}} + \sum_{k} \hbar \Omega_{k} \parens{\hat{\sigma}_{g\uparrow,k} + \hat{\sigma}_{\uparrow g,k}}
\\
& + \sum_{k} \hbar g_{e,k} \parens{\hat{b}^{\dagger} \hat{\sigma}_{\downarrow\uparrow,k} + \hat{b} \hat{\sigma}_{\uparrow\downarrow,k}} + \hbar\eta_b\parens{\hat{b}^{\dagger}+\hat{b}}.
\end{split}
\end{equation}
where $\hat{a}$ and $\hat{a}^\dagger$ represent the annihilation and creation operator of the weak optical Fabry-Perot mode, $\omega_{o,c}$ is the optical mode frequency, and $\eta_b$ sets the strength of quantum noise fluctuations that perturb the microwave cavity field, which in turn randomly kicks the Bloch spin vectors away from $z$-axis.

Similar to the microwave-only case, we employ the Heisenberg equations of motion with a mean-field approximation to gain intuitive insight into the full three-level cavity QED dynamics of our system. The resulting equations of motion are given by:
\begin{align}
\dirac{\dot{\hat{a}}} &= -i\omega_{o,c}\dirac{\hat{a}}-\frac{\kappa_o}{2} \dirac{\hat{a}}-i\sum_k g_{o,k}\dirac{\hat{\sigma}_{g\downarrow,k}}
\\
\dirac{\dot{\hat{b}}} &= -i\omega_{e,c}\dirac{\hat{b}}-i\sum_k g_{e,k}\dirac{\hat{\sigma}_{23,k}}-\frac{\kappa_e}{2}\dirac{\hat{b}}-i\eta_b
\\
\dirac{\dot{\hat{\sigma}}_{g\downarrow,k}} &= -(i\omega_{g\downarrow}+\frac{\gamma_o}{2})\dirac{\hat{\sigma}_{g\downarrow,k}}+i\Omega_k\dirac{\hat{\sigma}_{\uparrow\downarrow,k}}-ig_{o,k}(\dirac{\hat{\sigma}_{gg,k}}-\dirac{\hat{\sigma}_{\downarrow\downarrow,k}}) \dirac{\hat{a}}-ig_{e,k}\dirac{\hat{b}^{\dagger}}\dirac{\hat{\sigma}_{g\uparrow,k}}
\\
\dirac{\dot{\hat{\sigma}}_{\downarrow\uparrow,k}} &= -i(\omega_{\downarrow\uparrow}+\frac{\gamma_s}{2})\dirac{\hat{\sigma}_{\downarrow\uparrow,k}}-ig_{e,k}(\dirac{\hat{\sigma}_{\downarrow\downarrow,k}}-\dirac{\hat{\sigma}_{\uparrow\uparrow,k}})\dirac{\hat{b}}+ig_{o,k}\dirac{\hat{a}^{\dagger}}\dirac{\hat{\sigma}_{g\uparrow,k}}-i\Omega_k\dirac{\hat{\sigma}_{\downarrow g,k}}
\\
\dirac{\dot{\hat{\sigma}}_{g\uparrow,k}} &= -(i\omega_{g\uparrow}+\frac{\gamma_o}{2})\dirac{\hat{\sigma}_{g\uparrow,k}}-ig_{e,k}\dirac{\hat{\sigma}_{g\downarrow,k}}\dirac{\hat{b}}-i\Omega_k(\dirac{\hat{\sigma}_{gg,k}}-\dirac{\hat{\sigma}_{\uparrow\uparrow,k}})+ig_{o,k}\dirac{\hat{\sigma}_{\downarrow\uparrow,k}}\dirac{\hat{a}}
\\
\dirac{\dot{\hat{\sigma}}_{gg,k}} &= \gamma_1 \dirac{\hat{\sigma}_{\downarrow\downarrow,k}} +\gamma_1 \dirac{\hat{\sigma}_{\uparrow\uparrow,k}} - i\Omega(\dirac{\hat{\sigma}_{g\uparrow,k}}-\dirac{\hat{\sigma}_{\uparrow g,k}}) - ig_{o,k}(\dirac{\hat{a}^{\dagger}}\dirac{\hat{\sigma}_{g\downarrow,k}} - \dirac{\hat{\sigma}_{\downarrow g,k}}\dirac{\hat{a}}) 
\\
\dirac{\dot{\hat{\sigma}}_{\downarrow \downarrow,k}} &= -\gamma_1 
\dirac{\hat{\sigma}_{\downarrow \downarrow,k}} - ig_{e,k}(\dirac{\hat{b}^{\dagger}}\dirac{\hat{\sigma}_{\downarrow \uparrow,k}} - \dirac{\hat{\sigma}_{\uparrow \downarrow,k}}\dirac{\hat{b}}) - ig_{o,k}(\dirac{\hat{\sigma}_{\downarrow g1,k}}\dirac{\hat{a}}-\dirac{\hat{a}^{\dagger}}\dirac{\hat{\sigma}_{g \downarrow,k}})
\\
\dirac{\dot{\hat{\sigma}}_{\uparrow\uparrow,k}} &= -\gamma_1 \dirac{\hat{\sigma}_{\uparrow \uparrow,k}} - i\Omega(\dirac{\hat{\sigma}_{\uparrow g,k}}-\dirac{\hat{\sigma}_{g \uparrow,k}})-ig_{e,k}(\dirac{\hat{\sigma}_{\uparrow\downarrow,k}}\dirac{\hat{b}}-\dirac{\hat{b}^{\dagger}}\dirac{\hat{\sigma}_{\downarrow\uparrow,k}})
\label{Simu_3levels_eqn}
\end{align}
Here, cavity decay terms as well as the relaxation and dephasing terms for the atomic operators are included, where $\kappa_o$ is the optical cavity decay rate, $\kappa_e$ is the microwave cavity decay rate, $\gamma_o$ is the optical dephasing rate, $\gamma_s$ is the spin dephasing rate, and $\gamma_1$ is the population decay rate.

\begin{figure}[h]
\includegraphics[width=1\linewidth]{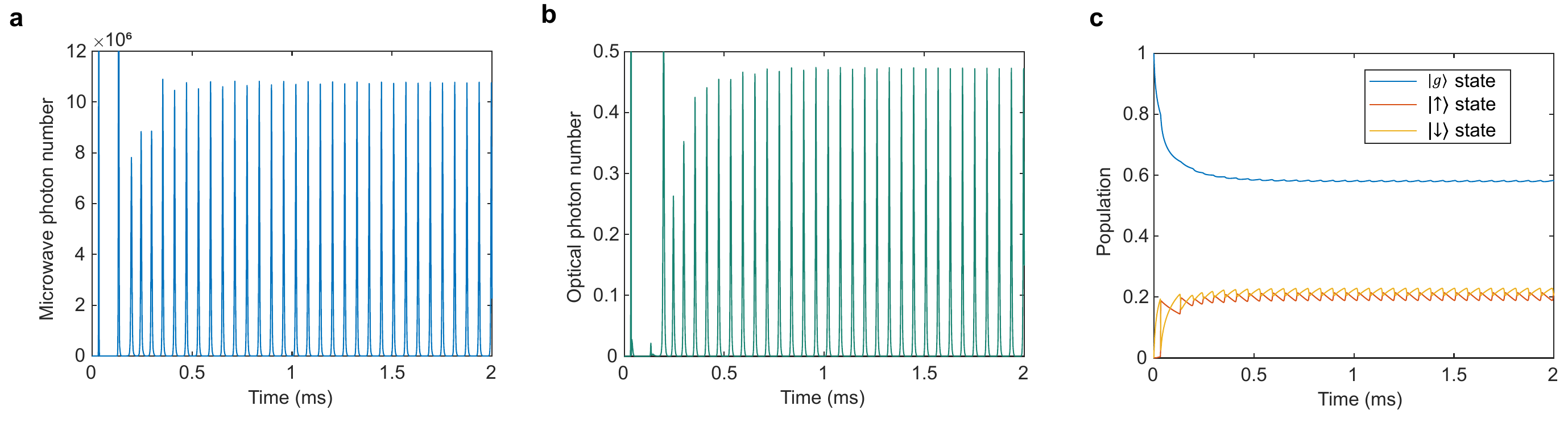}
\caption{\textbf{Numerical simulation of periodic superradiance via a mean-field method.} (a) Calculated microwave field obtained from the mean-field equations. (b) Calculated optical field showing the emergence of pulsed periodic superradiance, transferred from the microwave superradiant pulse trains. (c) Calculated populations of the three levels, $\{ \ket{g}, \ket{\uparrow}, \ket{\downarrow} \}$, averaged over the entire spin ensemble. We find that the mean-field simulation yields a periodicity of $\approx$60~$\mu$s, which is in discrepancy with the measured $\approx$80~$\mu$s, as the mean-field approximation neglects spin–spin correlations.
}\label{3level-periodic-simu}
\end{figure}

The above equations establish a framework for simulating the cavity QED dynamics of a disordered spin ensemble within the mean-field approximation. Specifically, we prepare a disordered spin ensemble by sampling one hundred atoms from the spin inhomogeneous linewidth. Then, by solving the equations of motion using our experimental parameters corresponding to a pulsed superradiance regime (Regime III), we numerically observe the emergence of dual-rail periodic superradiance in both optical and microwave domains, as shown in Fig.~\ref{3level-periodic-simu}. Notice that the optical emission pattern follows the microwave emission pattern, as both originate from the same persistent spin dynamics. We find the mean-field simulation yields a periodicity of $\approx$60 $\mu$s, compared to the measured $\approx$80 $\mu$s, as the mean-field approximation ignores the spin-spin correlations by factoring $\langle \hat{S}_{\uparrow\downarrow} \hat{S}_{\downarrow\uparrow} \rangle$ into $\langle \hat{S}_{\uparrow\downarrow} \rangle \langle \hat{S}_{\downarrow\uparrow} \rangle$.

In Regime II, where the CW superradiance is expected, we observe significant line narrowing below the inhomogeneous linewidth due to collectively enhanced coherence. Here, the carrier frequency of the CW superradiant emission is determined by solving the above equations in the frequency domain, which yields
\begin{align}
        \left[i\parens{\omega-\omega_e}-\frac{\kappa}{2}\right]\langle\hat{b}\rangle & - ig\langle\hat{S}_{\downarrow\uparrow}\rangle = 0\\
        \left[i\parens{\omega-\omega_k}-\frac{\kappa_s}{2}\right]\langle\hat{\sigma}_{\downarrow\uparrow,k}\rangle & -ig\langle\hat{\sigma}_{z,k}\rangle\langle\hat{b}\rangle = 0\\
        -D\parens{\langle\hat{\sigma}_{z,k}\rangle+1}+\gamma_1\parens{1-\langle\hat{\sigma}_{z,k}\rangle}&-2ig\parens{\langle\hat{b}^\dagger\rangle\langle\hat{\sigma}_{\downarrow\uparrow,k}\rangle-\langle\hat{\sigma}_{\uparrow\downarrow,k}\rangle\langle\hat{b}\rangle}=0.
\end{align}
Here, $\kappa_s = \gamma+\gamma_1+D$. Solving these equations together yields
\begin{align}
    \langle\hat{\sigma}_{z,k}\rangle &= \frac{[i\parens{\omega-\omega_c}-\frac{\kappa}{2}][i\parens{\omega-\omega_k}-\frac{\kappa_s}{2}]}{-g^2}\frac{\langle\hat{\sigma}_{\downarrow\uparrow,k}\rangle}{\langle\hat{S}_{\downarrow\uparrow}\rangle}\\
    \langle \hat{S}_z\rangle &= \frac{1}{D+\gamma_1}\parens{\parens{\gamma_1-D}\frac{N}{2}+\frac{g^2\kappa\langle\hat{S}_{\uparrow\downarrow}\rangle\langle\hat{S}_{\downarrow\uparrow}\rangle}{\parens{\omega-\omega_e}^2+\parens{\kappa/2}^2}}.
\end{align}
Since $\langle \hat{\sigma}_{z,k} \rangle$ must be real, imposing $\mathrm{Im}\left[\langle \hat{\sigma}_{z,k} \rangle\right] = 0$ yields constraints on $\omega$ and $\langle \hat{\sigma}_{\downarrow\uparrow,k} \rangle$, which must be solved self-consistently to predict the CW emission frequency. As an example, for identical atoms where all $\omega_k$ are equal to a single frequency $\omega_s$, one obtains
\begin{equation}
    \omega = \frac{\kappa_s\omega_c+\kappa\omega_s}{\kappa+\kappa_s}
    \label{eq:emission_freq}
\end{equation}
which represents a weighted average of the cavity and spin frequencies, indicating a \textit{suppressed} cavity-pulling effect in the bad-cavity limit. 

\subsection{Efficient numerical simulation of a disordered spin ensemble}
\begin{figure}[h!]
\centering
\includegraphics[width=0.45\linewidth]{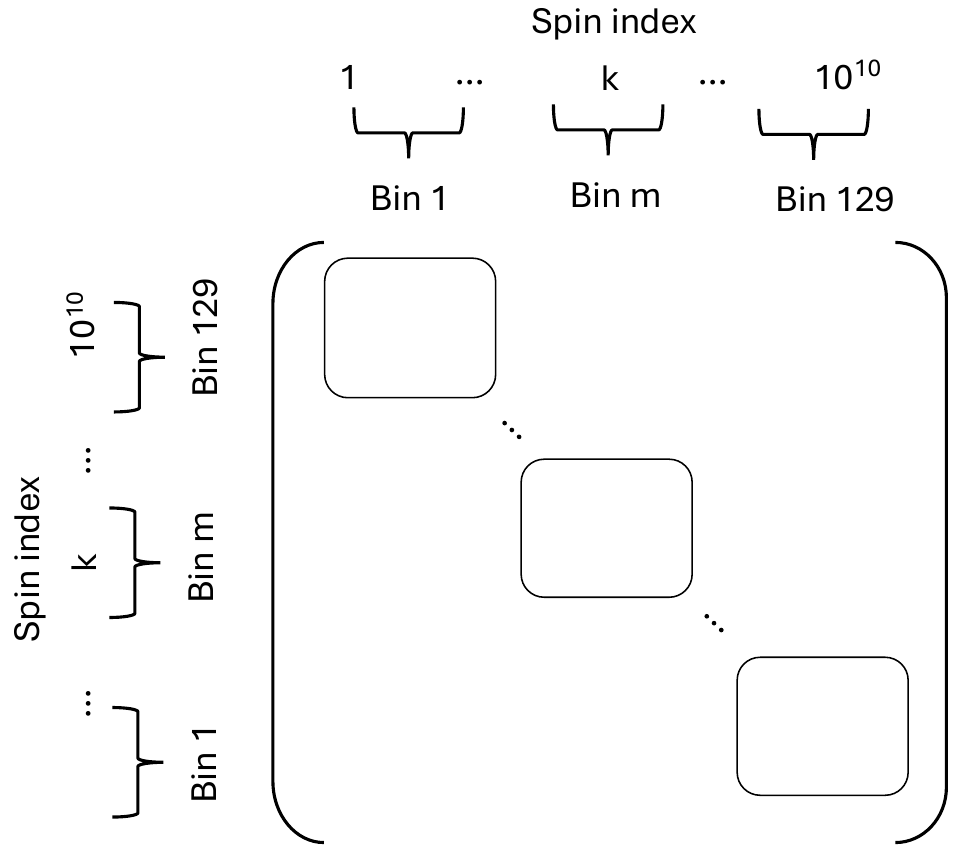}
\caption{
\textbf{Illustration of frequency binning for efficient numerical simulation.} The spin inhomogeneous linewidth is discretized into $M = 129$ bins, where $N = 10^{10}$ spins are sampled from a Gaussian distribution and sorted into these bins, which allows us to significantly reduce the effective size of the problem.}
\label{SI_binning}
\end{figure}

Directly simulating the second-order cumulant equations shown in Methods for $N \approx 10^{10}$ spins is computationally prohibitive. To enable efficient simulation, we sample $N = 10^{10}$ atoms from a Gaussian distribution representing the inhomogeneous spin-ensemble linewidth and group them into $M = 129$ frequency bins, treating atoms within each bin as identical. Given the experimentally measured inhomogeneous linewidth of $\approx$160 kHz, each bin has a width of $\approx$1.6 kHz, which is well below the estimated homogeneous linewidth of $\approx$16 kHz~\citeSI{bartholomew2020chip}, thereby justifying this approximation. This approach reduces the problem to $(M^2 + 2M + 1) \approx 10^4$ equations, as shown in Fig.~\ref{SI_binning}. 

The $m$-th frequency bin ($m = 1, 2, \cdots, M$) contains $N \rho_m$ spins, where $m$ indexes the bins and $\rho_m$ denotes the sampling density derived from the original inhomogeneous ensemble distribution. Using this frequency-binning scheme, we rewrite the second-cumulant expansion equations in the binned representation:
\begin{align}
    \frac{d}{dt}\langle \hat{b}^\dagger\hat{\sigma}_{\downarrow\uparrow,m}\rangle &= \parens{i\parens{\omega_e-\omega_{m}}-\frac{\kappa+\kappa_s}{2}} \langle \hat{b}^\dagger\hat{\sigma}_{\downarrow\uparrow,m}\rangle \nonumber\\
    &+ ig \parens{\frac{1-\langle \hat{\sigma}_{z,m}\rangle}{2}+(N \rho_m-1)\langle \hat{\sigma}_{\uparrow\downarrow,m}\hat{\sigma}_{\downarrow\uparrow,m}\rangle+\sum_{n \neq m}N\rho_n \langle\hat{\sigma}_{\uparrow\downarrow,n}\hat{\sigma}_{\downarrow\uparrow,m}\rangle-\langle \hat{\sigma}_{z,m}\rangle\langle \hat{b}^\dagger \hat{b}\rangle}\\
    \frac{d}{dt}\langle \hat{b}^\dagger \hat{b}\rangle &= -\kappa \langle \hat{b}^\dagger \hat{b}\rangle+\kappa N_\text{th}+ig\parens{\sum_m N \rho_m \langle \hat{\sigma}_{\uparrow\downarrow,m}\hat{b}\rangle-\sum_m N \rho_m\langle \hat{b}^\dagger \hat{\sigma}_{\downarrow\uparrow,m}\rangle}\\
    \frac{d}{dt}\langle \hat{\sigma}_{\uparrow\downarrow,m}\hat{\sigma}_{\downarrow\uparrow,n}\rangle &= \parens{i\omega_m-i\omega_n-\kappa_s}\langle \hat{\sigma}_{\uparrow\downarrow,m}\hat{\sigma}_{\downarrow\uparrow,n}\rangle +ig\langle \hat{\sigma}_{z,m}\rangle\langle \hat{b}^\dagger \hat{\sigma}_{\downarrow\uparrow,n}\rangle-ig\langle \hat{\sigma}_{z,n}\rangle\langle \hat{\sigma}_{\uparrow\downarrow,m}\hat{b}\rangle\\
    \frac{d}{dt}\langle \hat{\sigma}_{z,m}\rangle &= -D\parens{\langle \hat{\sigma}_{z,m}\rangle+1}+\gamma_1\parens{1-\langle \hat{\sigma}_{z,m}\rangle}-2ig\parens{\langle \hat{b}^\dagger \hat{\sigma}_{\downarrow\uparrow,m}\rangle-\langle \hat{\sigma}_{\uparrow\downarrow,m}\hat{b}\rangle}.
\end{align}

Note that $\langle \hat{S}_{\uparrow\downarrow}\hat{S}_{\downarrow\uparrow}\rangle = \sum_k\sum_{j\neq k}\langle \hat{\sigma}_{\uparrow\downarrow,j}\hat{\sigma}_{\downarrow\uparrow,k}\rangle$, where $j$ and $k$ index individual spins. This expression can be rewritten in the frequency-binned representation as
\begin{align}
    \langle \hat{S}_{\uparrow\downarrow}\hat{S}_{\downarrow\uparrow}\rangle = \sum_m N \rho_m (N \rho_m-1)\langle \hat{\sigma}_{\uparrow\downarrow,m}\hat{\sigma}_{\downarrow\uparrow,m}\rangle+\sum_m N \rho_m \sum_{n \neq m}N\rho_n \langle\hat{\sigma}_{\uparrow\downarrow,n}\hat{\sigma}_{\downarrow\uparrow,m}\rangle,
\end{align}
where $m$ and $n$ index the frequency bins. Therefore, the spin–spin correlation can be computed more efficiently by representing the collective spin operator as a $129 \times 129$ matrix in the frequency-binned representation, thereby significantly reducing the effective dimensionality of the problem.

\begin{figure}[h]
\centering
\includegraphics[width=0.8\linewidth]{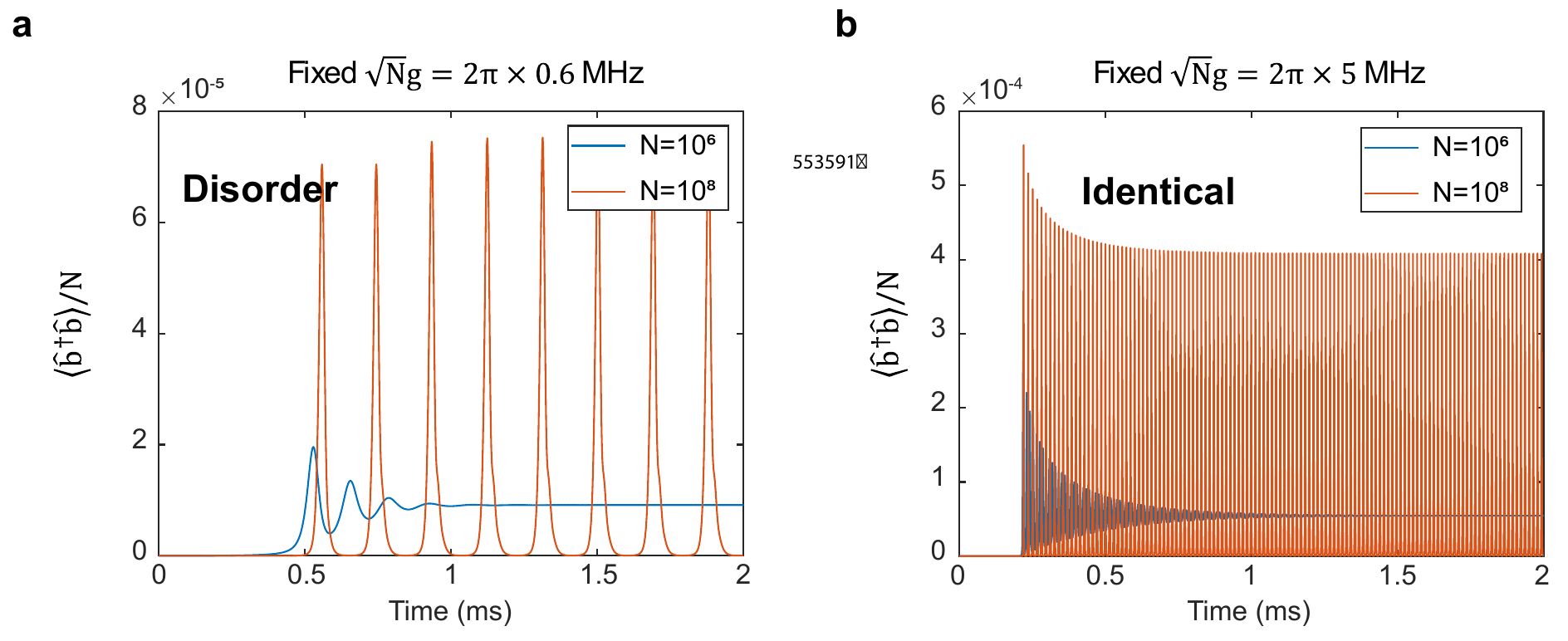}
\caption{\textbf{Numerical simulations illustrating the emergence of the pulsed superradiance phase as a collective effect.} (a) Comparison of emission patterns for two different ensemble sizes with disorder, $N = 10^6$ (blue) and $N = 10^8$ (red), with the same effective coupling strength, $\sqrt{N}g = 2\pi \times 0.6~\mathrm{MHz}$. (b) Comparison of emission patterns for two different ensemble sizes without disorder, $N = 10^6$ (blue) and $N = 10^8$ (red), with the same effective coupling strength, $\sqrt{N}g = 2\pi \times 5~\mathrm{MHz}$.  
}\label{2nd-N-threshold}
\end{figure}

Based on the above equations, together with the frequency-binning scheme, we numerically simulate the cavity-QED dynamics of a large-scale disordered spin ensemble, as shown in Extended Data Fig. 6. First, we confirm that the spin-spin correlation is more than an order of magnitude stronger than the stimulated emission, dominating the emission channel in the periodic superradiance regime. (Extended Data Fig. 6a). Second, we confirm that pulsed periodic superradiance originates from a large atomic ensemble: a single-spin model with an artificially increased cavity coupling strength, chosen to match that of a many-spin ensemble, fails to generate a pulse train in the emission (Extended Data Fig. 6b). Third, we confirm that the coupling threshold for the transition to the periodic superradiant regime is reduced in the presence of disorder (Extended Data Fig. 6c). In other words, the periodic superradiant phase can be obtained with a normalized coupling strength below unity (i.e., $g_{\mathrm{norm}} < 1$), whereas an identical homogeneous spin ensemble requires a normalized coupling strength greater than unity. From Extended Data Fig. 6b, it is evident that the pulsed superradiance phase is a collective phenomenon requiring a large number of emitters. To further investigate the dependence on the emitter number, $N$, we continuously vary $N$ while keeping the total coupling strength, $g_{\mathrm{tot}} = \sqrt{N}g$, fixed. As shown in Fig.~\ref{2nd-N-threshold}, we find that there exists a threshold in the minimum value of $N$ required to observe pulsed superradiance. A more detailed analytical derivation of this threshold will be presented in a later section.  

\begin{figure}[h]
\centering
\includegraphics[width=0.9\linewidth]{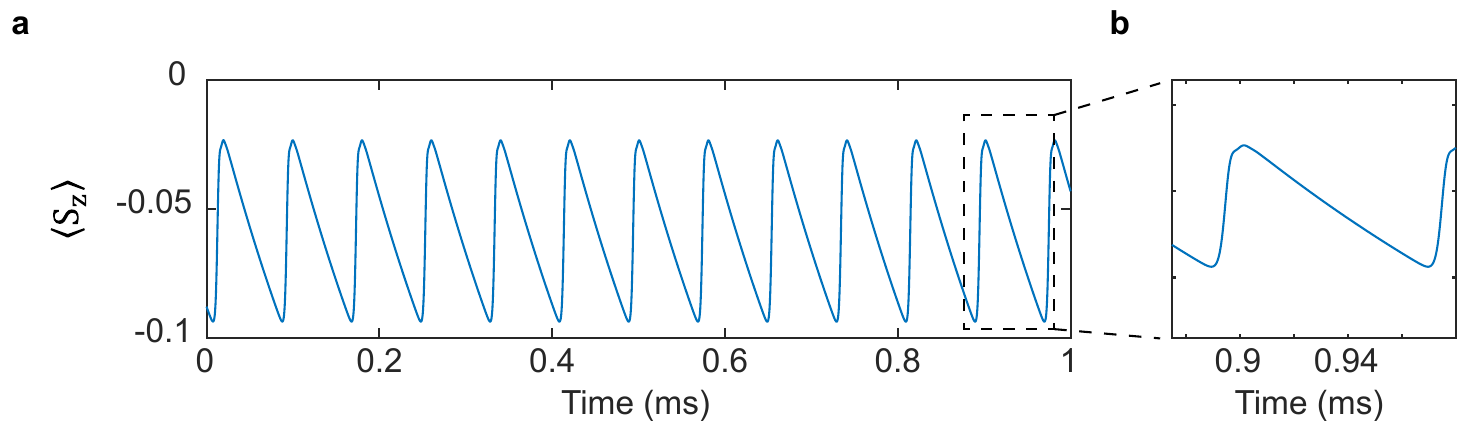}
\caption{\textbf{Numerical simulations illustrating the dynamics of the collective spin polarization.} (a) Evolution of $\langle \hat{S}_z \rangle$ in the periodic superradiance phase. (b) Zoomed-in view of the optical pumping process, showing linear growth of the population inversion followed by rapid decay due to a superradiant burst.
}\label{2nd-Sz}
\end{figure}

The emergence of a superradiant pulsed train requires the spin system to accumulate sufficient collective population inversion to initiate limit-cycle dynamics, as discussed in the main text. The dynamics of the collective spin polarization, $\langle \hat{S}_z \rangle$, are shown in Fig.~\ref{2nd-Sz}, exhibiting periodic behavior in which a nearly linear population accumulation buildup is observed due to optical pumping. 

\begin{figure}
\centering
\includegraphics[width=1\linewidth]{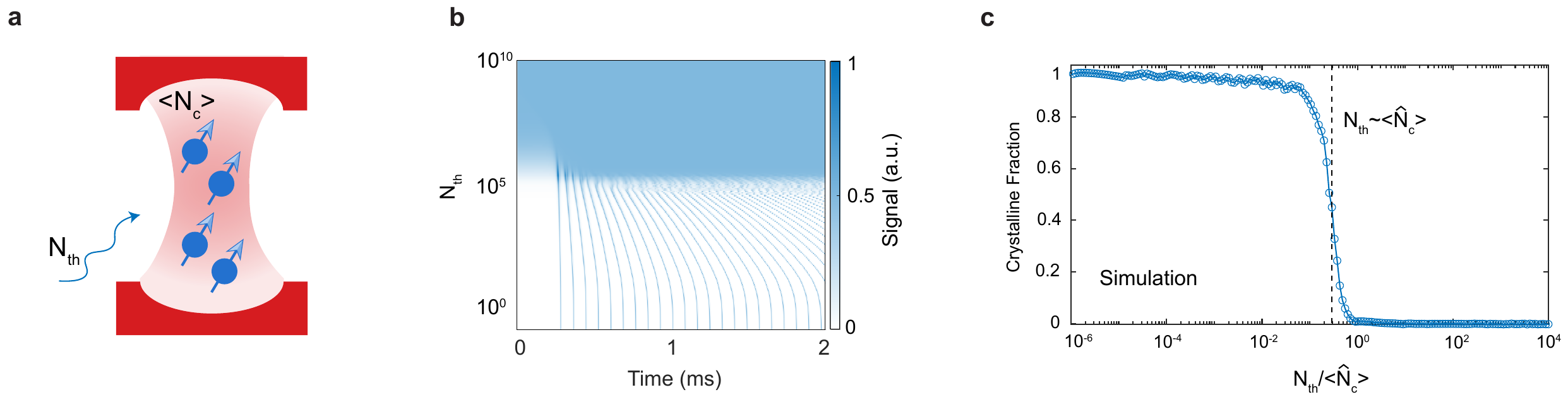}
\caption{
\textbf{Numerical simulation of the robustness of the periodic superradiant phase.} \textbf{a}, Thermal noise photons, $N_\text{th}$, can perturb the mean cavity photon number, $\langle \hat{N}_c \rangle.$ 
\textbf{b}, Numerically simulated pulsed superradiance time traces at various levels of $N_\text{th}$. Each time trace is normalized to its own maximum for visualization. \textbf{c}, Crystalline fraction as a function of the normalized resonator thermal occupancy defined as $N_\text{th}/\langle \hat{N}_c \rangle$. When the thermal noise is on the same order as the average intracavity photon number (dashed line), the periodic emission pattern disappears.
}
\label{SI_robustness}
\end{figure}

In our numerical simulations, we incorporate resonator noise by introducing a noise term, $N_{\mathrm{th}}$, into the cavity population equation via the input-output formalism. In Fig.~\ref{SI_robustness}, to test the robustness of the pulsed superradiance phase in Regime III, we vary the noise strength by tuning $N_{\mathrm{th}}$ in the simulations and examine whether periodic superradiance can still survive. We find that, as long as the thermal noise level, $N_\text{th}$, remains below the average intracavity photon number, the periodic pulsed superradiance phase persists robustly over long times without decay.

% \newpage
\section{Phase boundary estimation based on bifurcation}

\subsection{Simplified cavity-QED model with two spin sub-ensembles}
Here, we consider a cavity-coupled ensemble of two-level spin systems to analytically investigate the emergence of a periodic superradiance phase in a cavity-QED system. We begin with the mean-field equations of motion in the rotating frame of the cavity, given by
\begin{equation}
\begin{aligned}
    \frac{d}{dt}\langle \hat{b} \rangle &= -\frac{\kappa}{2} \langle \hat{b} \rangle - i g \sum_k \langle \hat{\sigma}_{\downarrow\uparrow,k} \rangle \\
    \frac{d}{dt}\langle \hat{\sigma}_{\downarrow\uparrow,k} \rangle &= -i \delta_k \langle \hat{\sigma}_{\downarrow\uparrow,k} \rangle - \gamma_s \langle \hat{\sigma}_{\downarrow\uparrow,k} \rangle - i g \langle \hat{\sigma}_{z,k}\rangle \langle \hat{b} \rangle \\
    \frac{d}{dt}\langle \hat{\sigma}_{z,k} \rangle &= -(D-\gamma) - (D+\gamma) \langle \hat{\sigma}_{z,k} \rangle + 2 i g \left(\langle \hat{\sigma}_{\uparrow\downarrow,k} \rangle \langle \hat{b} \rangle - \langle \hat{\sigma}_{\downarrow\uparrow,k} \rangle \langle \hat{b}^\dagger \rangle\right).
\end{aligned}
\label{eq:MF_withcavity}
\end{equation}
Here, $\hat{\sigma}_{z,k}$ is the population inversion (Pauli-$z$) operator of the $k$-th spin, $\hat{\sigma}_{\uparrow\downarrow,k}$ and $\hat{\sigma}_{\downarrow\uparrow,k}$ are the raising and lowering operators, and $\hat{b}$ and $\hat{b}^\dagger$ are the annihilation and creation operators of the cavity mode, respectively. 

Intuitively, the physical origin of the periodic superradiance phase can be understood as arising from the competition between effective cavity-mediated dissipative interactions among the emitters, which favor phase synchronization, and frequency disorder, which instead promotes dephasing. Since solving the above equations directly is analytically challenging due to the random, inhomogeneous distribution of $\delta_k$, we instead consider a simplified toy model in which the ensemble of $N$ emitters is divided into two sub-ensembles, each containing $N/2$ emitters (assuming $N$ is even for simplicity). The two sub-ensembles are assumed to have detunings of $\pm \delta$, where $\delta$ serves as a proxy for the disorder strength. While this simplification is not \textit{a priori} justified, we remark that a similar approximation was employed to study the stability of driven-dissipative spin coherent states in the presence of inhomogeneous broadening, yielding qualitatively similar results to numerical simulations of the full mean-field equations~\citeSI{mok2025stability,patra2019driven}. We can then rewrite the above mean-field equations in terms of collective operators for the sub-ensembles,
\begin{align}
    \hat{S}_z &\equiv \frac{1}{2} \sum_{k} \hat{\sigma}_{z,k} = \hat{S}_{1,z} + \hat{S}_{2,z}, \\ 
    \hat{S}_- &\equiv \sum_k \hat{\sigma}_{\downarrow\uparrow,k} = \hat{S}_{1,-} + \hat{S}_{2,-}, \\
    \hat{S}_+ &\equiv \sum_k \hat{\sigma}_{\uparrow\downarrow,k} = \hat{S}_{1,+} + \hat{S}_{2,+},
\end{align}
where $\hat{S}_{j,z}$ and $\hat{S}_{j,\pm}$ are supported on the $j$-th sub-ensemble ($j = 1,2$), to obtain
\begin{align}
    \frac{d}{dt}\langle \hat{b} \rangle &= -\frac{\kappa}{2} \langle \hat{b} \rangle - i g \left(\langle \hat{S}_{1,-} \rangle + \langle \hat{S}_{2,-} \rangle \right) \\
    \frac{d}{dt}\langle \hat{S}_{1,-} \rangle &= -i\delta \langle \hat{S}_{1,-} \rangle - \gamma_s \langle \hat{S}_{1,-} \rangle - 2 i g \langle \hat{S}_{1,z} \rangle \langle \hat{b} \rangle \\
    \frac{d}{dt}\langle \hat{S}_{2,-} \rangle &= +i\delta \langle \hat{S}_{2,-} \rangle - \gamma_s \langle \hat{S}_{2,-} \rangle - 2 i g \langle \hat{S}_{2,z} \rangle \langle \hat{b} \rangle \\
    \frac{d}{dt}\langle \hat{S}_{1,z} \rangle &= -\frac{N}{4} \gamma_- - \gamma_+ \langle \hat{S}_{1,z} \rangle + i g \left(\langle \hat{S}_{1,+} \rangle \langle \hat{b} \rangle - \langle \hat{S}_{1,-} \rangle \langle \hat{b}^\dagger \rangle \right) \\
    \frac{d}{dt}\langle \hat{S}_{2,z} \rangle &= -\frac{N}{4} \gamma_- - \gamma_+ \langle \hat{S}_{2,z} \rangle + i g \left(\langle \hat{S}_{2,+} \rangle \langle \hat{b} \rangle - \langle \hat{S}_{2,-} \rangle \langle \hat{b}^\dagger \rangle \right).
    \label{eq:MF_twodetuned_subensemble}
\end{align}
We have introduced $\gamma_\pm \equiv D \pm \gamma$ for convenience. The difference between the sub-ensembles is captured by the variables
\begin{align}
    \hat{D}_z &\equiv \hat{S}_{1,z} - \hat{S}_{2,z}, \\
    \hat{D}_- &\equiv \hat{S}_{1,-} - \hat{S}_{2,-}, \\
    \hat{D}_- &\equiv \hat{S}_{1,+} - \hat{S}_{2,+}.
\end{align}
Note that, unlike $\hat{S}_\pm$ and $\hat{S}_z$, these difference variables are anti-symmetric under the exchange of the two sub-ensembles. With this, we can rewrite the original mean-field equations as
\begin{align}
    \frac{d}{dt}\langle \hat{b} \rangle &= -\frac{\kappa}{2} \langle \hat{b} \rangle - i g \langle \hat{S}_{-} \rangle \\
    \frac{d}{dt}\langle \hat{S}_{-} \rangle &= -i \delta \langle \hat{D}_{-} \rangle - \gamma_s \langle \hat{S}_{-} \rangle - 2 i g \langle \hat{S}_{z} \rangle \langle \hat{b} \rangle \\
    \frac{d}{dt}\langle \hat{D}_{-} \rangle &= -i \delta \langle \hat{S}_{-} \rangle - \gamma_s \langle \hat{D}_{-} \rangle - 2 i g \langle \hat{D}_{z} \rangle \langle \hat{b} \rangle \\
    \frac{d}{dt}\langle \hat{S}_{z} \rangle &= -\frac{N}{2} \gamma_- - \gamma_+ \langle \hat{S}_{z} \rangle + i g \left(\langle \hat{S}_{+} \rangle \langle \hat{b} \rangle - \langle \hat{S}_{-} \rangle \langle \hat{b}^\dagger \rangle\right) \\
    \frac{d}{dt}\langle \hat{D}_{z} \rangle &= -\gamma_+ \langle \hat{D}_{z} \rangle + i g \left(\langle \hat{D}_{+} \rangle \langle \hat{b} \rangle - \langle \hat{D}_{-} \rangle \langle \hat{b}^\dagger \rangle\right).
    \label{eq:MF_twodetuned}
\end{align}
The above equations illustrate the effect of the disorder $\delta$, which couples the symmetric variables $\{\hat{S}_\pm, \hat{S}_z\}$ to the anti-symmetric variables $\{\hat{D}_\pm, \hat{D}_z\}$. 

To simplify the analysis further, we assume that the system is initialized with $\langle \hat{S}_- \rangle \neq 0$, and $\langle \hat{D}_- \rangle = \langle \hat{D}_z \rangle = \langle \hat{b} \rangle = 0$. This corresponds to an initial symmetric state with non-zero coherence. The coherence can be chosen to be very small, which can still trigger the superradiant emission. Without loss of generality, we set the initial $\langle \hat{S}_-\rangle$ to be real. With this initial condition, the dynamics ensures that $\langle \hat{b} \rangle$ and $\langle \hat{D}_- \rangle$ are purely imaginary, while $\langle \hat{S}_z\rangle$ and $\langle \hat{S}_-\rangle$ are purely real, and $\langle \hat{D}_z \rangle = 0$. This motivates us to define the real variables
\begin{equation}
    w \equiv \frac{i \langle \hat{b} \rangle}{\sqrt{N}}, \quad x \equiv \frac{\langle \hat{S}_- \rangle}{N}, \quad y \equiv \frac{i\langle \hat{D}_-\rangle}{N}, \quad z \equiv -\frac{2\langle \hat{S}_z \rangle}{N}.
\end{equation}
Then, the above set of equations can be expressed as
\begin{align}
    \dot{w} &= -\frac{\kappa}{2} w + \tilde{g} x \\
    \dot{x} &= -\delta y - \gamma_s x + \tilde{g} w z\\
    \dot{y} &= +\delta x - \gamma_s y \\
    \dot{z} &= \gamma_- - \gamma_+ z - 4 \tilde{g} w x,
\label{eq:simplified_wxyz}
\end{align}
where $\tilde{g} \equiv \sqrt{N}g$. We are interested in the steady-state behavior, which can be obtained by setting $\dot{w} = \dot{x} = \dot{y} = \dot{z} = 0$ and solving the resulting equations. This yields two steady-state solutions, which we denote as $(w_0,x_0,y_0,z_0)$: The trivial solution
\begin{equation}
    w_0 = x_0 = y_0 = 0, \quad z_0 = \frac{\gamma_-}{\gamma_+},
\label{eq:trivial_ss}
\end{equation}
and the non-trivial solution
\begin{equation}
    w_0 = \frac{2\tilde{g}}{\kappa} x_0, \quad x_0 = \pm \sqrt{\frac{\kappa}{8\tilde{g}^2} (\gamma_- -  \gamma_+ z_0)}, \quad y_0 = \frac{\delta}{\gamma_s} x_0, \quad z_0 = \frac{\kappa }{2\tilde{g}^2 \gamma_s}(\gamma_s^2 + \delta^2).
\label{eq:nontrivial_ss}
\end{equation}
Physically, the trivial solution refers to a completely incoherent steady state with the cavity field in the vacuum state. On the other hand, the non-trivial solution describes a steady state in which the emitters have some coherence, with CW emission from the cavity. In the context of superradiant phase transitions, the trivial and non-trivial solutions are referred to as the normal and superradiant phases, respectively. We expect the system to undergo a transition from the normal to superradiant phase as the collective emitter-cavity coupling, $\tilde{g}$, is increased beyond some critical value, similar to the usual Dicke model without disorder (which corresponds to $\delta = 0$). 

Let us also comment on two features of the non-trivial solution. First, $y_0$, which measures the phase deviation between the two sub-ensembles, is proportional to $\delta/\gamma_s$. This means that, at least for small disorder strengths $\delta$, the superradiant phase is robust against disorder. The spins are approximately aligned along the same direction on the collective Bloch sphere, but with a small phase spread $\propto \delta/\gamma_s$. Second, the non-trivial solution has a twofold degeneracy. This stems from the $\mathbb{Z}_2$ symmetry under the transformation 
\begin{equation}
    w \to -w, \quad x \to -x, \quad y \to -y.
\end{equation}
Physically, this means that there is no preference for the spins to align on either one side or the opposite side of the Bloch sphere. The $\mathbb{Z}_2$ symmetry is also inherent in the full mean-field equations [Eq.~\eqref{eq:MF_withcavity}], under the transformation $\hat{\sigma}_{\downarrow\uparrow,k} \to -\hat{\sigma}_{\downarrow\uparrow,k}$ and $\hat{b} \to -\hat{b}$.

So far, this does not account for the observed periodic superradiance. To capture this, we need to analyze the stability of the steady-state solutions. We do this using linear stability analysis. Let us perturb the system about the steady state by introducing a small fluctuation $\vec{\epsilon} = (\epsilon_w,\epsilon_x,\epsilon_y,\epsilon_z)$. Linearizing the dynamics around the steady state, we obtain, to linear order in the fluctuation,
\begin{equation}
    \dot{\vec{\epsilon}} = \Lambda \vec{\epsilon},
\end{equation}
where $\Lambda$ is the Jacobian matrix, given by
\begin{equation}
    \Lambda = \begin{pmatrix}
        -\frac{\kappa}{2} && \tilde{g} && 0 && 0 \\
        \tilde{g}z_0 && -\gamma_s && -\delta && \tilde{g}w_0 \\
        0 && \delta && -\gamma_s && 0 \\
        -4\tilde{g}x_0 && -4\tilde{g} w_0 && 0 && -\gamma_+
    \end{pmatrix}.
\end{equation}
The eigenvalues of $\Lambda$ determine the linear stability of the steady state. In the following, we use this to analyze both the trivial and non-trivial solutions. To fix ideas, we also consider realistic parameters relevant to the experiment, as summarized in Extended Data Table 2.

\subsection{Stability analysis for the trivial solution}
Substituting the trivial solution [Eq.~\eqref{eq:trivial_ss}], the Jacobian matrix reads
\begin{equation}
    \Lambda = \begin{pmatrix}
        -\frac{\kappa}{2} && \tilde{g} && 0 && 0 \\
        \frac{\tilde{g}\gamma_-}{\gamma_+} && -\gamma_s && -\delta && 0 \\
        0 && \delta && -\gamma_s && 0 \\
        0 && 0 && 0 && -\gamma_+
    \end{pmatrix},
\end{equation}
which has a block-diagonal form. Let us denote the upper $3 \times 3$ block as $\Lambda^\prime$. It can be shown that
\begin{equation}
    tr (\Lambda^\prime) = -\frac{\kappa}{2} - 2 \gamma_s < 0
\end{equation}
and
\begin{equation}
    \det \Lambda^\prime = \frac{\tilde{g}^2 \gamma_s \gamma_-}{\gamma_+} - \frac{\kappa}{2}(\gamma_s^2 + \delta^2).
\end{equation}
A sufficient condition for the trivial steady state to be unstable is $tr(\Lambda^\prime) < 0$ and $\det \Lambda^\prime < 0$. Therefore,
\begin{equation}
    \gamma_s^2 + \delta^2 < \frac{2\tilde{g}^2 \gamma_s \gamma_-}{\kappa \gamma_+}
\label{eq:trivial_instability}
\end{equation}
guarantees the instability of the trivial solution. This means that the trivial solution does not occur if the dephasing ($\gamma_s$) or disorder ($\delta$) is sufficiently small, which is the regime corresponding to our experimental conditions.

\subsection{Stability analysis for the non-trivial solution}
Next, we analyze the stability of the non-trivial solution [Eq.~\eqref{eq:nontrivial_ss}], which physically corresponds to CW superradiant emission. It can be shown that the same condition [Eq.~\eqref{eq:trivial_instability}] is also necessary for the non-trivial solution to be physically valid, i.e., for $x_0^2 > 0$. Diagonalizing the Jacobian matrix in this case is analytically intractable. However, we can still write down the characteristic equation for $\Lambda$:
\begin{equation}
    \lambda^4 + c_3 \lambda^3 + c_2 \lambda^2 + c_1 \lambda + c_0 = 0,
\label{eq:charEq}
\end{equation}
where $\lambda$ denotes the eigenvalues or the roots of the characteristic equation, and the coefficients are given by
\begin{equation}
\begin{aligned}
    c_3 &= \gamma_+ + 2 \gamma_s + \frac{\kappa}{2} \\
    c_2 &= \delta^2 + \frac{2\tilde{g}^2 \gamma_-}{\kappa} + \frac{1}{2}\sparens{(\gamma_+ + \gamma_s)(2\gamma_s + \kappa) - \frac{\delta^2}{\gamma_s}(2\gamma_+ + \kappa)} \\
    c_1 &= \frac{2\tilde{g}^2 \gamma_- (\gamma_s + \kappa)}{\kappa} - \frac{\gamma_+ \kappa(\gamma_s^2 + 3\delta^2)}{2\gamma_s}\\
    c_0 &= 2\tilde{g}^2 \gamma_- \gamma_s - \gamma_+ \kappa (\gamma_s^2 + \delta^2).
\end{aligned}
\end{equation}

Before delving into the solutions of the characteristic equation, we first present an intuitive explanation for the physical origin of periodic superradiance. For small disorder strength $\delta$, the CW emission (i.e., the non-trivial solution) is stable, reflecting the robustness of the superradiant phase. As the disorder strength increases, a critical value $\delta_c$ is reached at which the non-trivial solution loses stability via a Hopf bifurcation~\citeSI{abraham1988dynamical}. Upon further increasing $\delta$ beyond the Hopf bifurcation point, a stable limit cycle emerges in the phase space around $(w_0,x_0,y_0,z_0)$, corresponding to sustained periodic dynamics. We argue that, for the parameters relevant to our experiment, the disorder strength lies beyond this bifurcation point, thereby accounting for the observed periodic superradiance.

To support our argument, we identify the critical disorder $\delta_c$ at which the Hopf bifurcation occurs. Recall that a Hopf bifurcation arises when a pair of complex-conjugate eigenvalues of $\Lambda$ crosses the imaginary axis at $\delta = \delta_c$. Therefore, at this point, the characteristic equation [Eq.~\eqref{eq:charEq}] takes the form
\begin{equation}
    (\lambda - \mu)(\lambda - \nu)(\lambda - i\Omega)(\lambda + i\Omega) = 0,
\label{eq:charEq_Hopf}
\end{equation}
where $\Lambda$ has a pair of purely imaginary eigenvalues $\lambda = \pm i\Omega$, and two additional eigenvalues $\mu$ and $\nu$, which may in general be complex. Comparing Eqs.~\eqref{eq:charEq} and~\eqref{eq:charEq_Hopf}, we obtain 
\begin{equation}
    c_3 = -(\mu + \nu), \quad c_2 = \mu \nu + \Omega^2, \quad c_1 = -(\mu + \nu)\Omega^2, \quad c_0 = \mu \nu \Omega^2.
\end{equation}
Imposing consistency among these relations yields
\begin{equation}
    \frac{c_0 c_3}{c_1} = c_2 - \frac{c_1}{c_3},
\label{eq:hopf_condition}
\end{equation}
where both sides are equal to $\mu \nu$. This condition defines the Hopf bifurcation point.

The phase diagram shown in Fig.~1 of the main text is therefore obtained from Eq.~(\ref{eq:hopf_condition}) by numerically evaluating it using our experimental parameters. Strictly speaking, the $y$-axis, which represents the normalized coupling strength, should be expressed as $\sqrt{(\gamma_-/\gamma_+)Ng}/(\kappa/2)$ as the spin relaxation time, $T_1$, modifies the steady-state population. In particular, the effective participating atom number is reduced to $(\gamma_-/\gamma_+)N$, which corresponds to the maximum population achievable under a pumping rate $D$. For simplicity, however, we omit the factor $\sqrt{\gamma_-/\gamma_+}$ in Fig.~1, noting that it is typically close to unity ($\approx$0.8 in our case). Meanwhile, the experimentally measured ensemble coupling strength under pumping already incorporates the effect of spin relaxation and therefore reflects the effective normalized coupling $g_{\mathrm{norm}}$, as shown in Fig.~2. 

The $C=1$ phase boundary is obtained from Eq.~(\ref{eq:nontrivial_ss}), where the condition $|x_0|\geq 0$ must be satisfied. The inhomogeneous linewidth is defined as $\frac{\Gamma}{2} = [\int_{-\infty}^{\infty}\frac{\rho(\delta)d\delta}{\frac{\gamma_0}{2}+i\delta}]^{-1}$, where $\rho(\delta)$ is the probability distribution function of the frequency detuning~\citeSI{kersten2023triggered} and $\gamma_0$ is the homogeneous linewidth. For the two sub-ensemble case, this yields $\Gamma = \gamma_0 \left(1 + \frac{4\delta^2}{\gamma_0^2}\right)$, which reduces the condition $x_0=0$ to $\frac{2Ng^2}{\kappa}\frac{\gamma_-}{\gamma_+}\frac{\gamma_s}{\gamma_s^2+\delta^2} = \frac{4Ng^2}{\kappa\Gamma}\frac{\gamma_-}{\gamma_+}\approx\frac{4Ng^2}{\kappa\Gamma}=C=1$, where we have $\gamma_s = \gamma_0/2$ based on our definition. In Fig.~1e of the main text, we define the normalized disorder strength as additional inhomogeneous contribution to spectral line broadening beyond the homogeneous linewidth: $\Gamma/\gamma_0 - 1 = \delta^2/\gamma_s^2$. Although Eq.~\eqref{eq:hopf_condition} can be solved analytically for the critical disorder $\delta_c$, the resulting expression is cumbersome. Instead, we solve it numerically using the experimental parameters in Extended Data Table 2, obtaining $\delta_c \approx 2\pi\times 5 \, \text{kHz}$, corresponding to a critical linewidth of $\Gamma_c = 2\pi\times 35$ kHz. Since $\Gamma = 2\pi\times160 \, \text{kHz} > \Gamma_c$ in the experiment, this supports our hypothesis that the observed periodic superradiance is driven by a Hopf bifurcation and arises from the emergence of a stable limit cycle. 

We emphasize that this analysis is based on a simplified toy model designed to capture the effects of inhomogeneous spectral broadening. Within this model, the required collective emitter-cavity coupling $\tilde{g}$ can be calculated for each disorder $\delta$. Consequently, for a fixed disorder strength $\delta$, reducing $\tilde{g}$ can transition the state back to the non-trivial steady state, resulting in a transition from periodic to CW superradiance. This behavior is consistent with our experimental observations, where CW emission is observed under weak pumping, while periodic emission emerges under strong pumping.

\subsection{Analysis of the many-body nature of periodic superradiance}
From the numerical simulations in Extended Data Fig. 6, we find that the periodic pulsed superradiant phase is intrinsically a many-body phenomenon that cannot be captured by a single spin with an artificially enhanced coupling strength. Here, we provide an analytical approach to understanding the critical ensemble size, $N$, required for the emergence of this dynamical phase in a superradiant many-body cavity-QED system. 

Specifically, we begin with the following master equation
\begin{equation}
    \dot{\hat{\rho}} = -i[\hat{H},\hat{\rho}] + \sum_{j=1}^{N} \parens{\gamma_\uparrow \mathcal{D}[\hat{\sigma}_{j,+}]\hat{\rho} + \gamma_{\downarrow} \mathcal{D}[\hat{\sigma}_{j,-}]\hat{\rho} + \gamma_\phi \mathcal{D}[\hat{\sigma}_{j,z}]\hat{\rho}} + \kappa \mathcal{D}[\hat{b}]\hat{\rho},
\end{equation}
where
\begin{equation}
    \hat{H} = \sum_{j=1}^{N} \frac{\delta_j}{2} \hat{\sigma}_{j,z} + g \sum_{j=1}^{N} \parens{\hat{\sigma}_{j,-} \hat{b}^\dagger + \hat{\sigma}_{j,+} \hat{b}}.
\end{equation}
For simplicity, we consider identical atoms without disorder ($\delta_j = 0$). Using the second-order cumulant expansion, the equations of motion read
\begin{align}
    \frac{d}{dt} \langle \hat{b}^\dagger \hat{b} \rangle &= 2 g \,\text{Im}\langle \hat{S}_{-} \hat{b}^\dagger \rangle - \kappa \langle \hat{b}^\dagger \hat{b} \rangle \\
    \frac{d}{dt} \,\text{Im}\langle \hat{S}_{-} \hat{b}^\dagger \rangle &= g \langle \hat{S}_{z} \rangle \langle \hat{b}^\dagger \hat{b} \rangle + g \parens{\langle \hat{C} \rangle + \frac{N + \langle \hat{S}_{z} \rangle}{2}} - \parens{\gamma_s + \frac{\kappa}{2}} \text{Im}\langle \hat{S}_{-} \hat{b}^\dagger \rangle \\
    \frac{d}{dt} \langle \hat{S}_{z} \rangle &= N \gamma_- - \gamma_+ \langle \hat{S}_{z} \rangle - 4 g \,\text{Im}\langle \hat{S}_{-} \hat{b}^\dagger \rangle \\
    \frac{d}{dt}\langle \hat{C} \rangle &= 2 g \langle \hat{S}_{z} \rangle \,\text{Im}\langle \hat{S}_{-} \hat{b}^\dagger \rangle - 2\gamma_s \langle \hat{C} \rangle.
\end{align}
In the above equations, we defined $\gamma_\pm \equiv \gamma_\uparrow \pm \gamma_\downarrow$, the total spin dephasing rate $\gamma_s = \frac{1}{2}\parens{\gamma_\uparrow + \gamma_\downarrow + 4\gamma_\phi}$, and the total spin-spin correlation operator $\hat{C} = \sum_{\substack{{j,k=1} \\ {j\neq k}}}^{N} \hat{\sigma}_{\uparrow\downarrow,j} \hat{\sigma}_{\downarrow\uparrow,k}.$ 

In particular, we assume the following scaling with $N$:
\begin{equation}
    \langle \hat{b}^\dagger \hat{b} \rangle \sim N, \quad
    g \sim \frac{1}{\sqrt{N}}, \quad
    \text{Im}\langle \hat{S}_{-} \hat{b}^\dagger \rangle \sim N^{3/2}, \quad
    \langle \hat{S}_{z} \rangle \sim N, \quad
    \langle \hat{C} \rangle \sim N^2,
\end{equation}
with all other parameters of order unity. This motivates the introduction of the rescaled variables
\begin{equation}
    n = \frac{\langle \hat{b}^\dagger \hat{b} \rangle}{N}, \quad
    x = \frac{\text{Im}\langle \hat{S}_{-} \hat{b}^\dagger \rangle}{N^{3/2}}, \quad
    z = \frac{\langle \hat{S}_{z} \rangle}{N}, \quad
    c = \frac{\langle \hat{C} \rangle}{N^2},
\end{equation}
along with $\tilde{g} = \sqrt{N} g$. In terms of these variables, the equations of motion become
\begin{equation}
\begin{aligned}
    \dot{n} &= 2\tilde{g} x - \kappa n \\
    \dot{x} &= \tilde{g}\parens{z n + c + \frac{1+z}{2N}} - \parens{\gamma_s+\frac{\kappa}{2}} x \\
    \dot{z} &= \gamma_- - \gamma_+ z - 4\tilde{g}x \\
    \dot{c} &= 2\tilde{g} z x - 2\gamma_s c
\end{aligned}
\end{equation}
For sufficiently large $N$, we expect $z n + c \gg \frac{1+z}{2N}$. Setting $\dot{n} = \dot{x} = \dot{z} = \dot{c} = 0$ to obtain the steady-state solution, we find 
\begin{equation}
\begin{aligned}
    0 &= 2\tilde{g} x - \kappa n \\
    0 &= \tilde{g}\parens{z n + c} - \parens{\gamma_s + \frac{\kappa}{2}} x \\
   0 &= \gamma_- - \gamma_+ z - 4\tilde{g}x \\
    0 &= 2\tilde{g} z x - 2\gamma_s c
\end{aligned}
\end{equation}
Eliminating $z$ from these equations yields a relation among the remaining variables:
\begin{equation}
    n = \frac{2\tilde{g}}{\kappa} x = \frac{4\tilde{g}^2}{\kappa^2} c.
\end{equation}
This physically implies that the cavity photon number ($n$), spin-cavity ($\tilde{g}$), and spin-spin (c) correlations are all proportional to one another. Solving the steady-state equations yields two possible solutions: The trivial solution
\begin{equation}
    n = x = c = 0, \quad z = \frac{\gamma_-}{\gamma_+},
\end{equation}
and the non-trivial solution
\begin{equation}
\begin{aligned}
    n &= \frac{\gamma_-}{2\kappa} \parens{1 - \frac{\gamma_+}{\gamma_-} z}, \\
    x &= \frac{\gamma_-}{2\sqrt{\Gamma \kappa}} \parens{1 - \frac{\gamma_+}{\gamma_-} z}, \\
    c &= \frac{\gamma_-}{2\Gamma} \parens{1 - \frac{\gamma_+}{\gamma_-} z}, \\
    z &= \frac{2\gamma_s}{\Gamma}.
\end{aligned}
\end{equation}
This predicts a steady-state value $z > 0$ and describes the asymptotic behavior in the large-$N$ limit. 

To estimate the system size at which this asymptotic behavior emerges, we substitute the solution back into our approximation. Self-consistency requires
\begin{equation}
    c \gg \frac{1+z}{2N}, \quad z n \gg \frac{1+z}{2N}
\end{equation}
which is equivalent to
\begin{equation}
    N \gg \frac{\Gamma}{\gamma_-} \frac{1 + z}{1 - \frac{\gamma_+}{\gamma_-} z}, \quad N \gg \frac{\Gamma \kappa}{2\gamma_s \gamma_-} \frac{1+z}{1-\frac{\gamma_+}{\gamma_-} z}.
\end{equation}
In the regime where $\kappa, \tilde{g} \gg \gamma_\pm, \gamma_\phi, z \ll 1$, and $\kappa \gg 2\gamma_s$, the above condition simplifies to
\begin{equation}
    N \gg \frac{\Gamma \kappa}{2 \gamma_s \gamma_-}. 
\end{equation}
For example, taking $\kappa = 2\pi \times 4 \,\text{MHz}$, $\tilde{g} = 2\pi \times 1 \, \text{MHz}$, $\gamma_\downarrow = 2\pi \times 0.44 \, \text{kHz}$, $\gamma_\uparrow = 2\pi \times 0.76 \, \text{kHz}$, $\gamma_s = 2\pi \times 16 \,\text{kHz}$, we find that $N \gg 4\times 10^4$ serves as the ensemble-size threshold for the onset of asymptotic behavior, with $z \approx 0.06$. This is in qualitative agreement with our numerical simulations.

\bibliographystyleSI{naturemag}
\bibliographySI{SINotes}

\end{document}